\documentclass{article}
\usepackage{geometry}
\usepackage[utf8]{inputenc}
\usepackage{graphicx}
\usepackage{amsmath}
\usepackage{amssymb}
\usepackage{setspace}
\usepackage{enumerate}
\usepackage[ruled,linesnumbered,vlined]{algorithm2e}
\usepackage{color}
\usepackage{array}
\newcolumntype{C}[1]{>{\centering\let\newline\\\arraybackslash\hspace{0pt}}m{#1}}
\usepackage{microtype}
\usepackage{hyperref}
\usepackage{bm}
\usepackage{cite}
\usepackage{multirow}
\usepackage{lineno}
\usepackage{nicefrac}

\newtheorem{defn}{\noindent $\mathbf{Definition}$}[section]

\newtheorem{theorem}[defn]{$\mathbf{Theorem}$}

\title{Fast ellipsoidal conformal and quasi-conformal parameterization of genus-0 closed surfaces}

\author{Gary P. T. Choi\thanks{Department of Mathematics, The Chinese University of Hong Kong
  ({ptchoi@math.cuhk.edu.hk}).}}

\date{}

\begin{document}

\maketitle
\begin{abstract}
Surface parameterization plays a fundamental role in many science and engineering problems. In particular, as genus-0 closed surfaces are topologically equivalent to a sphere, many spherical parameterization methods have been developed over the past few decades. However, in practice, mapping a genus-0 closed surface onto a sphere may result in a large distortion due to their geometric difference. In this work, we propose a new framework for computing ellipsoidal conformal and quasi-conformal parameterizations of genus-0 closed surfaces, in which the target parameter domain is an ellipsoid instead of a sphere. By combining simple conformal transformations with different types of quasi-conformal mappings, we can easily achieve a large variety of ellipsoidal parameterizations with their bijectivity guaranteed by quasi-conformal theory. Numerical experiments are presented to demonstrate the effectiveness of the proposed framework. 
\end{abstract}

\section{Introduction}
Surface parameterization, the process of mapping a complicated surface to a simpler domain, is one of the most fundamental tasks in computer graphics, geometry processing, and shape analysis. To obtain a meaningful and useful parameterization, it is common to consider minimizing certain distortion criteria in the mapping computation. Among different types of parameterizations, conformal parameterizations are widely used as they preserve angles and hence the local geometry. Quasi-conformal parameterizations, another class of surface parameterizations, have been increasingly popular in recent years because of their greater flexibility in handling not only conformal distortions but also other prescribed constraints. Also, for surfaces with different topologies, different target parameter domains are used. In particular, as genus-0 closed surfaces are topologically equivalent to the sphere, it is natural to consider using the unit sphere as the target parameter domain. However, for surfaces with more complex geometry such as an elongated shape, using the sphere as the parameter domain may induce a large geometric distortion and hence hinder the use of the parameterization in practical applications.

\begin{figure}[t]
    \centering
    \includegraphics[width=\textwidth]{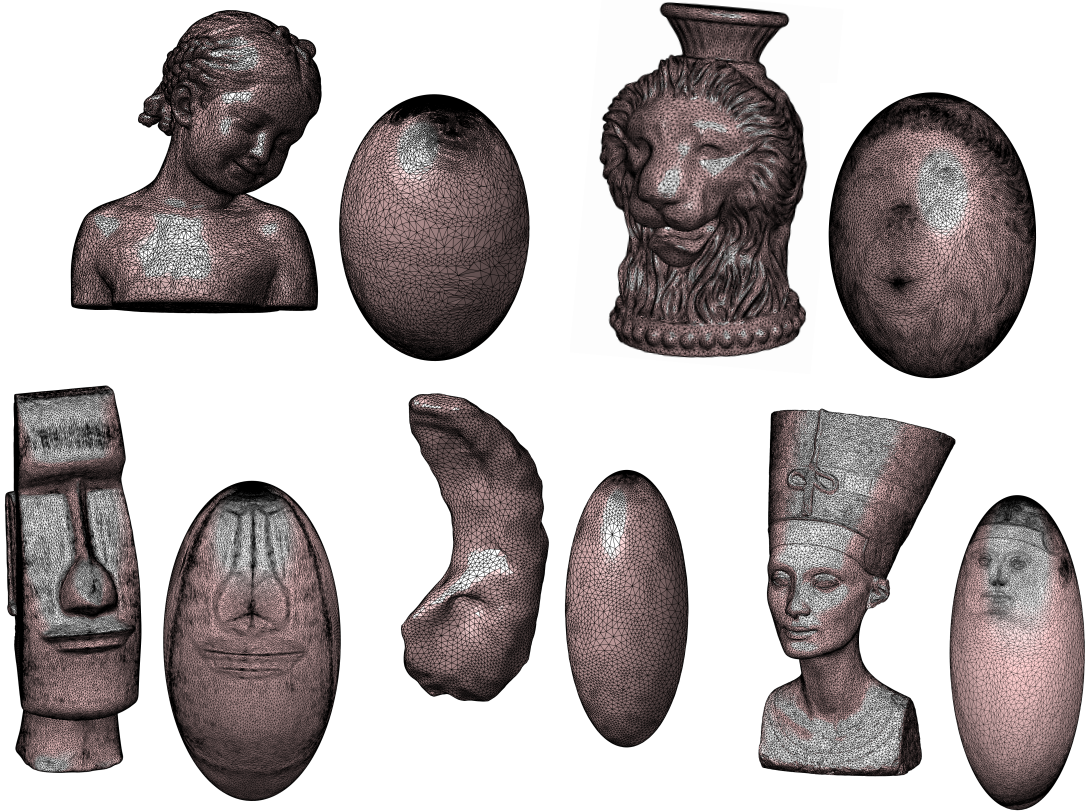}
    \caption{Ellipsoidal conformal parameterizations obtained by our proposed framework for different genus-0 closed surfaces. }
    \label{fig:illustration_fecm}
\end{figure}

In this paper, we tackle the problem of computing conformal and quasi-conformal parameterizations of genus-0 closed surfaces onto an \emph{ellipsoid} instead of a sphere. Specifically, we develop a framework that utilizes a composition of various conformal and quasi-conformal mappings to efficiently compute ellipsoidal conformal parameterizations for different genus-0 closed surfaces (see Fig.~\ref{fig:illustration_fecm} for examples). Also, using quasi-conformal theory, we can ensure the bijectivity of the parameterization. With the aid of our ellipsoidal parameterization method, the overall geometric distortion of the conformal parameterization of genus-0 closed surfaces can be significantly reduced when compared to the conventional spherical parameterization methods. We then further develop a method for computing ellipsoidal quasi-conformal parameterization of genus-0 closed surfaces with prescribed landmark constraints. Altogether, our work provides a set of efficient ellipsoidal parameterization tools for different genus-0 closed surfaces, which can then be utilized in various applications such as texture mapping, surface registration, surface remeshing, and shape analysis.

The organization of this paper is as follows. In Section~\ref{sect:related}, we review the previous works related to our problem. In Section~\ref{sect:background}, we introduce the concepts of conformal and quasi-conformal maps. In Section~\ref{sect:main}, we develop a fast and accurate method for computing ellipsoidal conformal parameterizations for genus-0 closed surfaces. In Section~\ref{sect:main_qc}, we extend the proposed method for computing ellipsoidal quasi-conformal parameterizations with prescribed landmark constraints. In Section~\ref{sect:experiments}, numerical experiments are presented to demonstrate the effectiveness of our proposed methods. We conclude our paper and discuss possible future works in Section~\ref{sect:conclusion}.

\section{Related works}\label{sect:related}
Over the past several decades, many surface parameterization methods have been developed. We refer the readers to~\cite{floater2005surface,sheffer2006mesh,choi2022recent} for comprehensive surveys. Below, we briefly review existing methods that are most relevant to our work.

For genus-0 closed surfaces, the most commonly used target parameter domain is the unit sphere $\mathbb{S}^2$. Therefore, numerous works have been devoted to the development of spherical parameterization methods. For spherical conformal parameterization, Angenent et al.~\cite{angenent1999laplace,haker2000conformal} proposed a linearization method for computing spherical conformal mappings. In~\cite{gu2002computing}, Gu and Yau developed a method for global conformal parameterization based on Hodge theory. Gu et al.~\cite{gu2004genus} developed a harmonic energy minimization method for computing spherical conformal mappings. In~\cite{sheffer2004robust}, Sheffer et al. developed a spherical angle-based flattening method for spherical conformal parameterization. Kharevych et al.~\cite{kharevych2006discrete} proposed a method for spherical conformal parameterization using circle patterns. In~\cite{springborn2008conformal}, Springborn et al. computed spherical conformal parameterization using discrete conformal equivalence. Later, several flow-based methods were developed for spherical conformal parameterization, including the surface Ricci flow~\cite{jin2008discrete,chen2013ricci}, mean curvature flow~\cite{kazhdan2012can}, and Willmore flow~\cite{crane2013robust}. In~\cite{lai2014folding}, Lai et al. proposed a harmonic energy minimization approach for folding-free spherical conformal parameterization. Choi et al.~\cite{choi2015flash} proposed a fast method for computing spherical conformal parameterization using quasi-conformal theory. A variant of the method was then developed in~\cite{choi2016spherical}. In~\cite{yueh2017efficient,liao2022convergence}, Yueh et al. developed a conformal energy minimization approach for spherical conformal parameterization. More recently, a parallelizable spherical conformal parameterization method was developed using partial welding~\cite{choi2020parallelizable}. 

Besides the above-mentioned spherical conformal parameterization methods, there are also many existing approaches for spherical parameterizations based on other distortion criteria. For instance, Praun and Hoppe~\cite{praun2003spherical} developed a stretch-based method for spherical parameterization with applications to remeshing. Gotsman et al.~\cite{gotsman2003fundamentals} developed a spherical parameterization method by generalizing the barycentric coordinates, which was later improved by Saba et al.~\cite{saba2005practical}. In~\cite{zayer2006curvilinear}, Zayer et al. developed a spherical parameterization method using curvilinear coordinate system. Lui et al.~\cite{lui2007landmark,choi2015flash} developed a method for landmark-constrained spherical parameterizations. In~\cite{athanasiadis2012feature}, Athanasiadis et al. proposed a feature-preserving spherical parameterization method based on geometrically constrained optimization. In~\cite{wang2014rigid}, Wang et al. developed an as-rigid-as-possible method for spherical parameterization. Later, Nadeem et al.~\cite{nadeem2016spherical} proposed a spherical parameterization method balancing angle and area distortion. In~\cite{wang2016bijective}, Wang et al. proposed a method for computing bijective spherical parameterizations with low distortion. In~\cite{choi2016fast,jarvis20213d}, Choi et al. developed methods for computing spherical quasi-conformal parameterizations. In~\cite{aigerman2017spherical}, Aigerman et al. proposed an algorithm for spherical orbifold Tutte embeddings. In~\cite{wang2018novel}, Wang et al. developed a local/global approach for spherical parameterization.

In recent years, there has been an increasing interest in exploring new parameter domains for surface parameterization. For instance, the unit hemisphere was used as the parameter domain for closed human brain and skull surfaces~\cite{giri2020open}. Spherical cap domains have also been used for the conformal or area-preserving parameterization of stone microstructures~\cite{shaqfa2021spherical} and anatomical structures~\cite{choi2022adaptive}. Besides, Lin et al.~\cite{lin2023ellipsoidal} utilized the Jacobi projection for computing mappings of genus-0 closed surfaces onto an ellipsoid. More recently, Shaqfa and van Rees~\cite{shaqfa2023spheroidal} developed a method for parameterizing star-shaped genus-0 objects onto a spheroid.

\section{Mathematical Background} \label{sect:background}
In this section, we introduce some basic concepts in conformal and quasi-conformal theory. Readers are referred to~\cite{lehto1973quasiconformal,gardiner2000quasiconformal,ahlfors2006lectures,gu2008computational} for more detail.

\subsection{Conformal maps}
Let $\mathbb{C} = \{x+iy: x,y \in \mathbb{R}\}$ be the complex plane, where $i$ is the imaginary number with $i^2 = -1$, and let $\overline{\mathbb{C}} = \mathbb{C} \cup \{\infty\}$ be the extended complex plane. We can express a map $f:\overline{\mathbb{C}} \to \overline{\mathbb{C}}$ as $f(z) = f(x,y) = u(x,y) + i v(x,y)$, where $z = x+iy$, and $u(x,y), v(x,y)$ are two real-valued functions. Suppose the derivative of the map $f$ is non-zero everywhere. $f$ is said to be a \emph{conformal} map if it satisfies the Cauchy--Riemann equations:
\begin{equation}\label{eqt:cauchyriemann}
    \frac{\partial u}{\partial x} = \frac{\partial v}{\partial y}  \ \ \text{ and } \ \ \frac{\partial u}{\partial y} = -\frac{\partial v}{\partial x}.
\end{equation}
The above equations can be rewritten as 
\begin{equation}
    \frac{\partial f}{\partial \overline{z}} = 0,
\end{equation}
where 
\begin{equation}\label{eqt:fzbar}
    \frac{\partial f}{\partial \overline{z}} = f_{\overline{z}} = \frac{1}{2}\left(\frac{\partial f}{\partial x} + i\frac{\partial f}{\partial y}\right).
\end{equation}
Note that conformal maps preserve angles and hence the local shapes. However, size is not preserved under conformal maps in general. 

M\"obius transformations are a class of conformal mappings $h:\overline{\mathbb{C}} \to \overline{\mathbb{C}}$ on the extended complex plane in the following form:
\begin{linenomath*}
\begin{equation}\label{eqt:mobius}
    h(z) = \frac{az+b}{cz+d},
\end{equation}
\end{linenomath*}
with $a,b,c,d \in \overline{\mathbb{C}}$ and $ad-bc \neq 0$.

\begin{figure}[t]
    \centering
    \includegraphics[width=\textwidth]{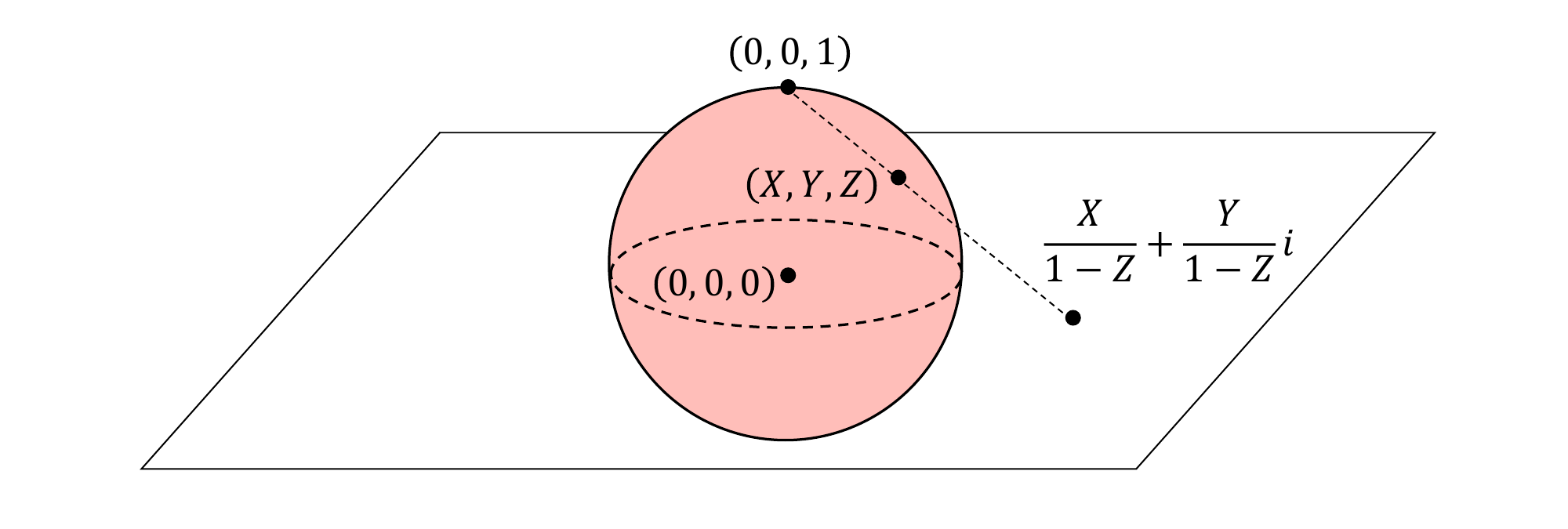}
    \caption{An illustration of the stereographic projection ${P^N}:\mathbb{S}^2 \to \overline{\mathbb{C}}$.}
    \label{fig:stereographic}
\end{figure}

Conformal maps can also be defined between Riemann surfaces. For any Riemann surfaces $\mathcal{M}$ and $\mathcal{N}$ with metrics $\mathbf{g}_{\mathcal{M}}$ and $\mathbf{g}_{\mathcal{N}}$, a diffeomorphism $f:\mathcal{M} \to {\mathcal{N}}$ is said to be conformal if the pull-back metric $f^*\mathbf{g}_{\mathcal{N}} = \lambda \mathbf{g}_{\mathcal{M}}$ for some positive function $\lambda$. In particular, the {(north-pole)} stereographic projection ${P^N}:\mathbb{S}^2 \to \overline{\mathbb{C}}$ is a conformal map from the unit sphere to the extended complex plane given by:
\begin{equation} \label{eqt:stereographic}
    {P^N}(X,Y,Z) = \frac{X}{1-Z}+\frac{Y}{1-Z}i,
\end{equation}
where $(X,Y,Z) \in \mathbb{S}^2$ (see Fig.~\ref{fig:stereographic} for an illustration). The inverse stereographic projection ${(P^N)}^{-1}:\overline{\mathbb{C}} \to \mathbb{S}^2$ is also a conformal map. For any $z = x+yi \in \overline{\mathbb{C}}$, we have
\begin{equation} \label{eqt:inverse_stereographic}
    {(P^N)}^{-1}(z) = {(P^N)}^{-1}(x+yi) = \left(\frac{2x}{1+x^2+y^2}, \frac{2y}{1+x^2+y^2}, \frac{-1+x^2+y^2}{1+x^2+y^2}\right).
\end{equation}
Similarly, the south-pole stereographic projection $P^S:\mathbb{S}^2 \to \overline{\mathbb{C}}$ and its inverse $(P^S)^{-1}:\overline{\mathbb{C}} \to \mathbb{S}^2$, given by
\begin{equation} \label{eqt:stereographic_south}
    P^S(X,Y,Z) = \frac{X}{1+Z}+\frac{Y}{1+Z}i,
\end{equation}
\begin{equation} \label{eqt:inverse_stereographic_south}
    (P^S)^{-1}(z) = (P^S)^{-1}(x+yi) = \left(\frac{2x}{1+x^2+y^2}, \frac{2y}{1+x^2+y^2}, \frac{1-x^2-y^2}{1+x^2+y^2}\right),
\end{equation}
are also conformal maps.

\subsection{Quasi-conformal maps}

Quasi-conformal maps are a generalization of conformal maps. Mathematically, let $f:\overline{\mathbb{C}} \to \overline{\mathbb{C}}$ be an orientation-preserving homeomorphism. $f$ is said to be a \emph{quasi-conformal} map if it satisfies the Beltrami equation:
\begin{equation}\label{eqt:beltrami}
\frac{\partial f}{\partial \overline{z}} = \mu_f(z) \frac{\partial f}{\partial z}
\end{equation}
for some complex-valued function $\mu_f$ with $||\mu_f||_{\infty}< 1$, where $f_{\overline{z}}$ is given by Eq.~\eqref{eqt:fzbar} and $f_z$ is given by the following equation:
\begin{equation}
    \frac{\partial f}{\partial z} = f_{z} =  \frac{1}{2}\left(\frac{\partial f}{\partial x} - i\frac{\partial f}{\partial y}\right).
\end{equation}

\begin{figure}[t]
    \centering
    \includegraphics[width=\textwidth]{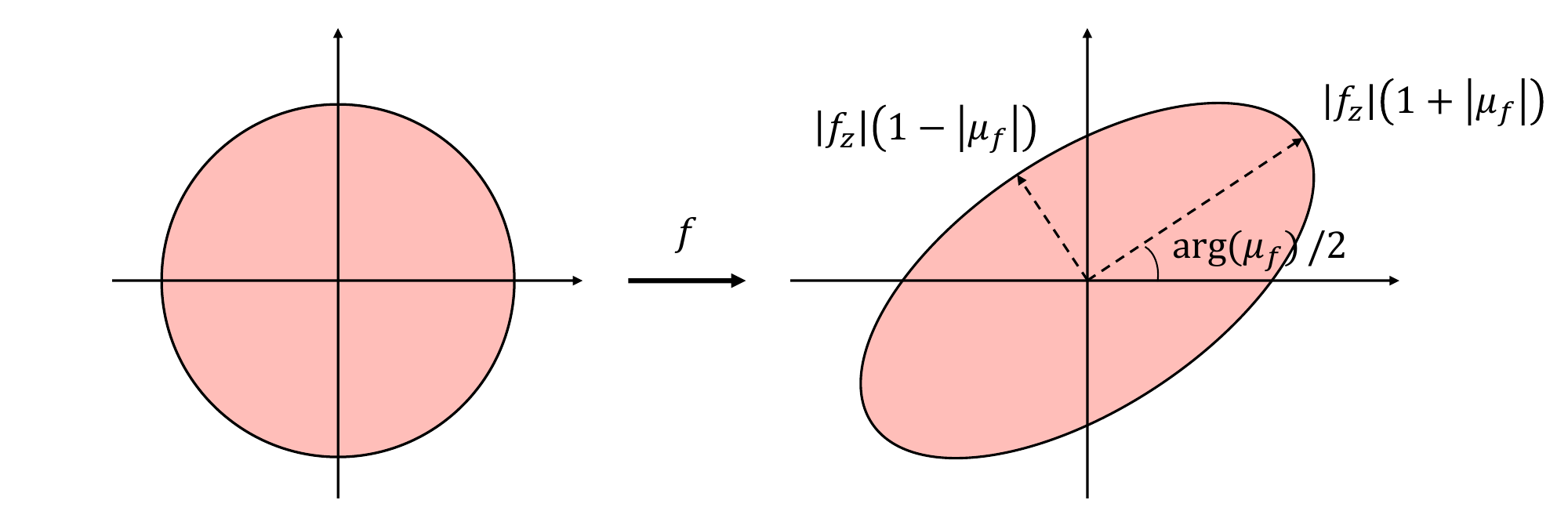}
    \caption{An illustration of quasi-conformal maps. Under a quasi-conformal map $f$, an infinitesimal circle is mapped to an infinitesimal ellipse with bounded eccentricity. The maximal magnification, maximal shrinkage, and orientation change are all related to the Beltrami coefficient $\mu_f$.}
    \label{fig:qc_figure}
\end{figure}

The complex-valued function $\mu_f$ is called the \emph{Beltrami coefficient} of the map $f$, which encodes important information about the quasi-conformal distortion of $f$. From Eq.~\eqref{eqt:cauchyriemann} and Eq.~\eqref{eqt:beltrami}, it is easy to see that $f$ is conformal at a point $p$ if and only if $\mu_f(p) = 0$. Intuitively, conformal mappings map infinitesimal circles to infinitesimal circles, while quasi-conformal maps map infinitesimal circles to infinitesimal ellipses with bounded eccentricity. More specifically, the maximal magnification factor and maximal shrinkage factor at a point $p$ are given by $\left|f_z(p)\right|\left(1+|\mu_f(p)|\right)$ and $\left|f_z(p)\right|\left(1-\left|\mu_f(p)\right|\right)$ respectively, and the orientation change of the major axis of the infinitesimal ellipse under $f$ is given by $\arg(\mu_f(p))/2$ (see Fig.~\ref{fig:qc_figure} for an illustration). The quasi-conformal dilatation of $f$ is given by
\begin{equation}
K(z) = \frac{1+|\mu_f(z)|}{1-|\mu_f(z)|}.
\end{equation}
Besides, the Jacobian $J_f$ of the mapping $f$ is given by
\begin{equation}
    J_f = |f_z|^2 \left(1-|\mu_f|\right)^2.
\end{equation}
Consequently, $J_f$ is positive everywhere if $\|\mu_f\|_{\infty} < 1$. This suggests that the bijectivity of a map is also related to its Beltrami coefficient. 

Moreover, by the Measurable Riemann Mapping Theorem~\cite{gardiner2000quasiconformal}, any given Beltrami coefficient can uniquely determine a quasi-conformal map under some suitable normalization. Computationally, the Linear Beltrami solver (LBS)~\cite{lui2013texture} can be used for efficiently reconstructing a quasi-conformal map from any given Beltrami coefficient. More specifically, let $\mu(z) = \rho(z) + i \tau(z)$ be the given Beltrami coefficient, where $\rho, \tau$ are real-valued functions. The corresponding quasi-conformal map $f(z) = f(x+iy) = u(x,y) + i v(x,y)$, where $u,v$ are real-valued functions, can be obtained by solving the following elliptic PDEs:
\begin{equation}\label{eqt:BeltramiPDE}
\left\{\begin{array}{cc}
    \nabla \cdot \left(A \nabla u \right) &= 0 \\
   \nabla \cdot \left(A \nabla v \right) &= 0,
\end{array}\right.
\end{equation}
where $A = \left( \begin{array}{cc}\alpha_1 & \alpha_2\\
\alpha_2 & \alpha_3 \end{array}\right)$ with 
\begin{equation}
    \alpha_1 = \frac{(\rho -1)^2 + \tau^2}{1-\rho^2 - \tau^2}, \ \ \ \alpha_2 = -\frac{2\tau}{1-\rho^2 - \tau^2}, \ \ \ \alpha_3 = \frac{1+2\rho+\rho^2 +\tau^2}{1-\rho^2 - \tau^2}.
\end{equation}
In the discrete case, the above elliptic PDEs can be discretized as sparse positive definite linear systems and hence can be solved efficiently (see~\cite{lui2013texture} for more details).

A useful property of the composition of quasi-conformal maps is as follows. Suppose $f:\Omega_1 \subset \mathbb{C} \to \Omega_2 \subset \mathbb{C}$ and $g:\Omega_2 \subset \mathbb{C} \to \Omega_3 \subset \mathbb{C}$ are two quasi-conformal maps. Then, the Beltrami coefficient of the composition $g \circ f$ is given by the following composition formula:
\begin{equation}\label{eqt:composition}
    \mu_{g \circ f} = \frac{\mu_f + \frac{\overline{f_{z}}}{f_z} (\mu_g \circ f)}{1+\frac{\overline{f_{z}}}{f_z} \overline{\mu_f} (\mu_g \circ f)}.
\end{equation}

One can also define quasi-conformal maps between Riemann surfaces. For any Riemann surfaces ${\mathcal{M}}$ and ${\mathcal{N}}$, an orientation-preserving diffeomorphism $f: {\mathcal{M}} \to {\mathcal{N}}$ is said to be quasi-conformal associated with the Beltrami differential $\mu(z)\frac{\overline{dz}}{dz}$ if for any chart $(U_{\alpha}, \phi_{\alpha})$ on ${\mathcal{M}}$ and any chart $(U_\beta, \psi_\beta)$ on ${\mathcal{N}}$, the composition map $f_{\alpha \beta} := \psi_{\beta} \circ f \circ {\phi_{\alpha}^{-1}}$ is quasi-conformal associated with $\mu_{\alpha} \frac{ \overline{dz_{\alpha}}}{dz_{\alpha}}$. Here, the Beltrami differential is an assignment to each chart $(U_{\alpha}, \phi_{\alpha})$ of an $L^{\infty}$ complex-valued function $\mu_{\alpha}(z_{\alpha})$ such that $\mu_{\alpha}(z_{\alpha}) \frac{ \overline{dz_\alpha}}{dz_\alpha} = \mu_{\beta}(z_{\beta}) \frac{ \overline{dz_{\beta}}}{dz_{\beta}}$ on the domain also covered by another chart $(U_\beta, \psi_\beta)$, where
$\frac{d z_{\beta}}{d z_{\alpha}} = \frac{d}{d z_{\alpha}} \phi_{\alpha \beta}$ and $\phi_{\alpha \beta} = \phi_{\beta} \circ \phi_{\alpha}^{-1}$.

\section{Fast ellipsoidal conformal map (FECM)}\label{sect:main}
Let $\mathcal{M}$ be a genus-0 closed surface discretized in the form of a triangle mesh $(\mathcal{V},\mathcal{F})$, where $\mathcal{V}$ is the vertex set and $\mathcal{F}$ is the triangulation. Define $\mathcal{E}_{a,b,c}$ as an ellipsoid with elliptic radii $a,b,c$:
\begin{linenomath*}
\begin{equation}
    \mathcal{E}_{a,b,c} = \left\{(x,y,z)\in \mathbb{R}^3: \frac{x^2}{a^2} + \frac{y^2}{b^2} + \frac{z^2}{c^2} = 1\right\},
\end{equation}
\end{linenomath*}
where $a,b,c >0$ are three positive scalars. Our goal is to obtain an ellipsoidal conformal parameterization $f: \mathcal{M} \to \mathcal{E}_{a,b,c}$.

We first recall the uniformization theorem:
\begin{theorem}[Uniformization theorem~\cite{poincare1908uniformisation}]
    Every simply connected Riemann surface is conformally equivalent to one of the following: 
\begin{itemize}
\item The open unit disk;
\item The complex plane;
\item The Riemann sphere.
\end{itemize}
\end{theorem}
In particular, every genus-0 closed surface is conformally equivalent to the unit sphere $\mathbb{S}^2$. Therefore, for any given genus-0 closed surface $\mathcal{M}$ and any given elliptic radii $a,b,c >0$, we can always find two conformal maps $f_1:\mathcal{M} \to \mathbb{S}^2$ and $f_2: \mathcal{E}_{a,b,c} \to \mathbb{S}^2$. Now, it is easy to see that the composition map $f_2^{-1} \circ f_1$ is a conformal map from $\mathcal{M}$ to $\mathcal{E}_{a,b,c}$. Consequently, we have the following theoretical guarantee for our ellipsoidal conformal parameterization framework:
\begin{theorem}[Existence of ellipsoidal conformal parameterization]
For any given genus-0 closed surface $\mathcal{M}$ and any given elliptic radii $a,b,c>0$, there exists an ellipsoidal conformal parameterization $f: \mathcal{M}\to \mathcal{E}_{a,b,c}$.
\end{theorem}

With the above theoretical guarantee, in this section we focus on the development of a fast and accurate method for computing ellipsoidal conformal parameterizations. Moreover, note that the above-mentioned composition map $f_2^{-1} \circ f_1$ is not the only possible conformal map between $\mathcal{M}$ and $\mathcal{E}_{a,b,c}$. Therefore, it is natural to search for an optimal ellipsoidal parameterization that further reduces certain geometric distortions in addition to being conformal. The procedure of our proposed ellipsoidal conformal parameterization algorithm is introduced in the following sections.

\subsection{Initial spherical conformal parameterization}
We start by applying the spherical conformal parameterization method in~\cite{choi2015flash} to map the given genus-0 closed surface $\mathcal{M}$ onto the unit sphere $\mathbb{S}^2$. More specifically, the method produces a spherical conformal mapping $\varphi:\mathcal{M} \to \mathbb{S}^2$ with $\varphi = (P^S)^{-1} \circ \eta \circ P^S \circ (P^N)^{-1} \circ \xi$, where $\xi:\mathcal{M} \to \overline{\mathbb{C}}$ is a harmonic flattening map of the surface onto the plane, $P^N$ is the north-pole stereographic projection given by Eq.~\eqref{eqt:stereographic}, $P^S$ is the south-pole stereographic projection given by Eq.~\eqref{eqt:stereographic_south}, and $\eta: \overline{\mathbb{C}} \to \overline{\mathbb{C}}$ is a suitably constructed quasi-conformal map (see~\cite{choi2015flash} for more details). In practice, note that numerical errors may be induced in the computation of the composition $P^S \circ (P^N)^{-1}$, as a large number of points may be mapped to a small region onto the sphere under $(P^N)^{-1}$ for complex geometries. To alleviate this issue, we obtain the following explicit formula for the composition $P^S \circ (P^N)^{-1}$ and directly use it in our computation:
\begin{equation}
\begin{split}
\left(P^S \circ (P^N)^{-1}\right)(z) &= \left(P^S \circ (P^N)^{-1}\right)(x+iy) \\
 &= \frac{\frac{2x}{1+x^2+y^2}}{1+\frac{-1+x^2+y^2}{1+x^2+y^2}} + i\frac{\frac{2y}{1+x^2+y^2}}{1+\frac{-1+x^2+y^2}{1+x^2+y^2}}
 = \frac{x}{x^2+y^2}+ i\frac{y}{x^2+y^2} = \frac{z}{|z|^2}.
 \end{split}
\end{equation}
This allows us to avoid unnecessary computations on the sphere at the intermediate steps of the spherical parameterization.

It is noteworthy that for optimal computational performance, the method in~\cite{choi2015flash} always searches for the most regular triangle element of $\mathcal{M}$ and maps it to the north pole in the spherical parameterization. Because of the symmetry of the sphere, this step will not lead to any bias in the spherical parameterization result. However, as the target ellipsoid $\mathcal{E}_{a,b,c}$ in our ellipsoidal parameterization problem is not rotationally symmetric in general, we cannot assume that the most regular triangle will always correspond to the north pole of the ellipsoid. Therefore, before moving on to the subsequent procedures, we need an extra step to correct the position of the two poles in the spherical parameterization.

To achieve this, we first apply the stereographic projection ${P^N}:\mathbb{S}^2 \to \overline{\mathbb{C}}$ using Eq.~\eqref{eqt:stereographic} to map the spherical parameterization result onto the plane. { In practice, we have $P^N \circ \varphi = P^N \circ (P^S)^{-1} \circ \eta \circ P^S \circ (P^N)^{-1} \circ \xi$, in which $P^N \circ (P^S)^{-1}$ can be explicitly expressed as
\begin{equation}
\begin{split}
\left(P^N \circ (P^S)^{-1}\right)(z) &= \left(P^N \circ (P^S)^{-1}\right)(x+iy) \\
 &= \frac{\frac{2x}{1+x^2+y^2}}{1-\frac{1-x^2-y^2}{1+x^2+y^2}} + i\frac{\frac{2y}{1+x^2+y^2}}{1-\frac{1-x^2-y^2}{1+x^2+y^2}}
 = \frac{x}{x^2+y^2}+ i\frac{y}{x^2+y^2} = \frac{z}{|z|^2}.
 \end{split}
\end{equation}
Again, this procedure allows us to avoid unnecessary computation on the sphere and reduce the numerical errors induced by the projections.} Next, we consider a M\"obius transformation $g:\overline{\mathbb{C}}\to\overline{\mathbb{C}}$ in the form 
\begin{equation} \label{eqt:mobius_alignment}
    g(z) = e^{i \theta}\frac{z-z_0}{z-z_1},
\end{equation}
where $z_0, z_1 \in \overline{\mathbb{C}}$ and $\theta \in [-\pi, \pi)$ are constants. Note that the map $g$ maps two prescribed points $z_0$ and $z_1$ on $\overline{\mathbb{C}}$ to $0$ and $\infty$ respectively. For our ellipsoidal parameterization problem, here we can choose $z_0$ and $z_1$ to be the two points that correspond to the two desired polar points on $\mathcal{M}$. Then, under the mapping $g$, the two points will be mapped to $0$ and $\infty$ respectively. Consequently, under the inverse stereographic projection ${(P^N)}^{-1}:\overline{\mathbb{C}} \to \mathbb{S}^2$, the two points will be mapped to the south pole and the north pole of the unit sphere respectively. The other parameter $\theta$ provides us with the flexibility of fixing the rotational degree of freedom of the ellipsoidal parameterization. In practice, we can set 
\begin{equation}
    \theta = -\text{arg}\left(\frac{z_2-z_0}{z_2-z_1}\right)
\end{equation}
for some point $z_2 \in \overline{\mathbb{C}}$ that corresponds to a point on $\mathcal{M}$ which is desired to be aligned with the positive $x$-axis in the resulting parameterization. Then, under the M\"obius transformation $g$, we will have $\text{arg}(g(z_2)) = 0$ and hence $g(z_2)$ lies on the positive real axis. Consequently, it will be mapped to a point on the positive $x$-axis under the inverse stereographic projection ${(P^N)}^{-1}$. 

Overall, we obtain a spherical conformal parameterization ${(P^N)}^{-1} \circ g \circ {P^N} \circ \varphi$. We remark that the subsequent steps involve projecting the spherical parameterization back to the extended complex plane and performing some further operations on the plane. Therefore, for simplicity, here we can just keep the mapping $(g\circ {P^N} \circ \varphi):\mathcal{M} \to \overline{\mathbb{C}}$ and ignore the last inverse stereographic projection step for now.

\subsection{Optimal ellipsoidal conformal parameterization}
After getting the mapping $(g\circ {P^N} \circ \varphi):\mathcal{M} \to \overline{\mathbb{C}}$, we consider mapping it to the target ellipsoid $\mathcal{E}_{a,b,c}$.

We first define the \emph{(north-pole) ellipsoidal stereographic projection} as a map $P_{a,b,c}^N:\mathcal{E}_{a,b,c} \to \overline{\mathbb{C}}$ from the ellipsoid $\mathcal{E}_{a,b,c}$ to the extended complex plane, with the explicit formula given by:
\begin{equation}
    P_{a,b,c}^N(X,Y,Z) = \frac{X/a}{1-Z/c}+\frac{Y/b}{1-Z/c}i,
\end{equation}
where $(X,Y,Z)$ is a point on $\mathcal{E}_{a,b,c}$. Note that $P_{a,b,c}^N$ maps $(\pm a,0,0)$, $(0,\pm b,0)$, $(0,0,c)$ and $(0,0,-c)$ to $\pm 1$, $\pm i$, $\infty$ and $0$ respectively, and so geometrically $P_{a,b,c}^N$ is not exactly a perspective projection of the ellipsoid through the north pole as in Fig.~\ref{fig:stereographic}. Nevertheless, it is easy to see that for $(a,b,c) = (1,1,1)$ we have $P_{a,b,c}^N = P^N$, which indicates that the ordinary stereographic projection is a special case of the ellipsoidal stereographic projection. Moreover, one can check that $P_{a,b,c}^N$ is a bijective mapping for any given $a,b,c > 0$. In fact, the \emph{inverse (north-pole) ellipsoidal stereographic projection} $(P^{N}_{a,b,c})^{-1}:\overline{\mathbb{C}} \to \mathcal{E}_{a,b,c}$ can be expressed as follows. For any $z = x+yi \in \overline{\mathbb{C}}$, we have
\begin{equation} \label{eqt:inverse_ellipsoidal_stereographic}
\begin{split}
    (P^{N}_{a,b,c})^{-1}(z) &= (P^{N}_{a,b,c})^{-1}(x+yi) \\
    &= \left(\frac{2ax}{1+x^2+y^2}, \frac{2by}{1+x^2+y^2}, \frac{c(-1+x^2+y^2)}{1+x^2+y^2}\right).
\end{split}
\end{equation}

Note that by applying the inverse ellipsoidal stereographic projection $(P^{N}_{a,b,c})^{-1}: \overline{\mathbb{C}} \to \mathcal{E}_{a,b,c}$, we can already get an ellipsoidal parameterization $\left((P^{N}_{a,b,c})^{-1}\circ g\circ {P^N} \circ \varphi\right):\mathcal{M} \to \mathcal{E}_{a,b,c}$. However, analogous to the issue in spherical parameterizations~\cite{choi2015flash}, the above-mentioned steps may lead to an uneven distribution of points on the ellipsoid. Therefore, we follow the idea in~\cite{choi2015flash} to improve the point distribution in the ellipsoidal parameterization via an extra step of rescaling the planar parameterization result.

To achieve this, we also need the notion of the ellipsoidal stereographic projection with respect to the south pole as well as the inverse of this projection. More specifically, we define the \emph{south-pole ellipsoidal stereographic projection} as a map $P_{a,b,c}^S:\mathcal{E}_{a,b,c} \to \overline{\mathbb{C}}$ from the ellipsoid $\mathcal{E}_{a,b,c}$ to the extended complex plane:
\begin{equation}
    P_{a,b,c}^S(X,Y,Z) = \frac{X/a}{1+Z/c}+\frac{Y/b}{1+Z/c}i,
\end{equation}
where $(X,Y,Z) \in \mathcal{E}_{a,b,c}$. Similarly, we define the \emph{inverse south-pole ellipsoidal stereographic projection} $(P^{S}_{a,b,c})^{-1}:\overline{\mathbb{C}} \to \mathcal{E}_{a,b,c}$ as follows: For any $z = x+yi \in \overline{\mathbb{C}}$, we have
\begin{equation}
\begin{split}
    (P^{S}_{a,b,c})^{-1}(z) &= (P^{S}_{a,b,c})^{-1}(x+yi) \\&= \left(\frac{2ax}{1+x^2+y^2}, \frac{2by}{1+x^2+y^2}, \frac{-c(-1+x^2+y^2)}{1+x^2+y^2}\right).
\end{split}
\end{equation}

Now, we establish an invariance result:
\begin{theorem}\label{invariance}
Let $T_1$ and $T_2$ be two triangles on $\overline{\mathbb{C}}$. The product of the perimeters of $T_1$ and $\left(P^S_{a,b,c} \circ (P^N_{a,b,c})^{-1}\right)(T_2)$ is invariant under any arbitrary scaling of $T_1$ and $T_2$.
\end{theorem}
\textbf{Proof.} Denote the vertices of $T_1$ and $T_2$ by $\{p_i\}_{i=1}^3$ and $ \{q_i\}_{i=1}^3$ respectively. For any $z = x+iy$, we have
\begin{equation}
\begin{split}
\left(P^S_{a,b,c} \circ (P^N_{a,b,c})^{-1}\right)(z) &= \left(P^S_{a,b,c} \circ (P^N_{a,b,c})^{-1}\right)(x+iy) \\
&= P^S_{a,b,c}\left(\frac{2ax}{1+x^2+y^2}, \frac{2by}{1+x^2+y^2}, \frac{c(-1+x^2+y^2)}{1+x^2+y^2}\right)\\
 &= \frac{\frac{2ax}{1+x^2+y^2} / a}{1+\frac{c(-1+x^2+y^2)}{1+x^2+y^2} / c} + i\frac{\frac{2by}{1+x^2+y^2}/b} {1+\frac{c(-1+x^2+y^2)}{1+x^2+y^2} / c}\\
 &= \frac{x}{x^2+y^2}+ i\frac{y}{x^2+y^2} = \frac{z}{|z|^2}.
 \end{split}
\end{equation}
Hence, we have
\begin{equation}
    \text{Perimeter}(T_1) = \sum_{1 \leq i < j \leq 3} |p_i-p_j|
\end{equation}
and
\begin{equation}
    \text{Perimeter}\left(\left(P^S_{a,b,c} \circ (P^N_{a,b,c})^{-1}\right)(T_2)\right) = \sum_{1 \leq i < j \leq 3} \left|\frac{q_i}{|q_i|^2}-\frac{q_j}{|q_j|^2}\right|.
\end{equation}
Now suppose $T_1$ and $T_2$ are scaled by an arbitrary factor $k$. We have
\begin{equation}
\begin{aligned}
&\text{Perimeter}(kT_1) \times \text{Perimeter}\left(\left(P^S_{a,b,c} \circ (P^N_{a,b,c})^{-1}\right)(kT_2)\right)\\
=& \left(\sum_{1 \leq i < j \leq 3} |kp_i-kp_j|\right)\left(\sum_{1 \leq i < j \leq 3} \left|\frac{kq_i}{|kq_i|^2}-\frac{kq_j}{|kq_j|^2}\right|\right)\\
=& \left(k\sum_{1 \leq i < j \leq 3} |p_i-p_j|\right)\left(\frac{1}{k}\sum_{1 \leq i < j \leq 3} \left|\frac{q_i}{|q_i|^2}-\frac{q_j}{|q_j|^2}\right|\right)\\
=& \left(\sum_{1 \leq i < j \leq 3} |p_i-p_j|\right)\left(\sum_{1 \leq i < j \leq 3} \left|\frac{q_i}{|q_i|^2}-\frac{q_j}{|q_j|^2}\right|\right)\\
=& \text{Perimeter}(T_1) \times \text{Perimeter}\left(\left(P^S_{a,b,c} \circ (P^N_{a,b,c})^{-1}\right)(T_2)\right).
\end{aligned}
\end{equation}
Hence, the product of the two perimeters is invariant. \hfill $\blacksquare$

We remark that using a similar argument, we can show that the product of $\text{Area}(T_1)$ and $\text{Area}\left(\left(P^S_{a,b,c} \circ (P^N_{a,b,c})^{-1}\right)(T_2)\right)$ is also invariant. 

Motivated by Theorem~\ref{invariance}, we consider improving the point distribution on the ellipsoidal parameterization by rescaling the planar parameterization $g\circ {P^N} \circ \varphi$ using an optimal parameter. Specifically, we consider the north pole triangle $T_{N}$ and the south pole triangle $T_{S}$ in $g\circ {P^N} \circ \varphi$, which correspond to the outermost triangle and the innermost triangle containing the origin. By Theorem~\ref{invariance}, the product of the perimeters of $T_N$ and $\left(P^S_{a,b,c} \circ (P^N_{a,b,c})^{-1}\right)(T_S)$ is invariant under arbitrary scaling of the two triangles. Now, we rescale the entire planar parameterization $g\circ {P^N} \circ \varphi$ by a factor 
\begin{equation} \label{eqt:k}
    k = \frac{\sqrt{\text{Perimeter}(T_{N})\times \text{Perimeter}(\left(P^S_{a,b,c} \circ (P^N_{a,b,c})^{-1}\right)(T_S))}}{\text{Perimeter}(T_{N})}.
\end{equation}
Then, $kT_{N}$ and $\left(P^S_{a,b,c} \circ (P^N_{a,b,c})^{-1}\right)(kT_{S})$ will have the same perimeter. We then apply the inverse north-pole ellipsoidal stereographic projection to the rescaled planar parameterization to obtain an ellipsoidal parameterization $(P^N_{a,b,c})^{-1} \circ k(g\circ {P^N} \circ \varphi):\mathcal{M} \to \mathcal{E}_{a,b,c}$. Because of the above rescaling step, the two polar triangles in the ellipsoidal parameterization will have similar sizes.

\subsection{Quasi-conformal composition}
Unlike the ordinary stereographic projection, the ellipsoidal stereographic projections $P^N_{a,b,c}$ and $P^S_{a,b,c}$ are not conformal in general. However, we can apply the idea of quasi-conformal composition to correct the distortion and achieve a conformal projection of the ellipsoid.

Specifically, we search for another quasi-conformal map $\psi: \overline{\mathbb{C}} \to \overline{\mathbb{C}}$ with the same Beltrami coefficient as $(P^{N}_{a,b,c})^{-1}$:
\begin{equation}\label{eqt:psi}
    \psi = \textbf{LBS}\left(\mu_{(P^{N}_{a,b,c})^{-1}}\right),
\end{equation}
where $\textbf{LBS}(\cdot)$ denotes the Linear Beltrami solver method in~\cite{lui2013texture} and $\mu_{(P^{N}_{a,b,c})^{-1}}$ denotes the Beltrami coefficient of $(P^{N}_{a,b,c})^{-1}$. Here, while the codomain of $(P^{N}_{a,b,c})^{-1}$ is in $\mathbb{R}^3$ instead of $\mathbb{R}^2$, we can still use the following metric-based formulation to compute the Beltrami coefficient $\mu_{(P^{N}_{a,b,c})^{-1}}$:
\begin{equation}\label{eqt:mu_metric}
\mu_{(P^{N}_{a,b,c})^{-1}} = \frac{E-G+2iF}{E+G+2\sqrt{EG-F^2}},
\end{equation}
where $\begin{pmatrix} E & F\\ F & G\end{pmatrix}$ is the first fundamental form. More explicitly, from Eq.~\eqref{eqt:inverse_ellipsoidal_stereographic}, we have
\begin{equation}
     \frac{\partial(P^{N}_{a,b,c})^{-1}}{\partial x} = \left(\frac{2a(-x^2+y^2+1)}{(1+x^2+y^2)^2}, \frac{-4bxy}{(1+x^2+y^2)^2}, \frac{4cx}{(1+x^2+y^2)^2}\right)
\end{equation}
and
\begin{equation}
     \frac{\partial(P^{N}_{a,b,c})^{-1}}{\partial y} = \left(\frac{-4axy}{(1+x^2+y^2)^2}, \frac{2b(x^2-y^2+1)}{(1+x^2+y^2)^2}, \frac{4cy}{(1+x^2+y^2)^2}\right),
\end{equation}
and hence
\begin{equation}\label{eqt:E}
\begin{split}
    E &= \left\langle \frac{\partial(P^{N}_{a,b,c})^{-1}}{\partial x}, \frac{\partial(P^{N}_{a,b,c})^{-1}}{\partial x} \right\rangle
    \\&= \frac{4a^2(-x^2+y^2+1)^2 + 16b^2x^2y^2 + 16c^2x^2}{(1+x^2+y^2)^4},
\end{split}
\end{equation}
\begin{equation}\label{eqt:F}
\begin{split}
    F &= \left\langle \frac{\partial(P^{N}_{a,b,c})^{-1}}{\partial x}, \frac{\partial(P^{N}_{a,b,c})^{-1}}{\partial y} \right\rangle \\&= \frac{-8a^2xy(-x^2+y^2+1)-8b^2xy(x^2-y^2+1) + 16c^2xy}{(1+x^2+y^2)^4},
\end{split}
\end{equation}
\begin{equation}\label{eqt:G}
\begin{split}
    G &= \left\langle \frac{\partial(P^{N}_{a,b,c})^{-1}}{\partial y}, \frac{\partial(P^{N}_{a,b,c})^{-1}}{\partial y} \right\rangle \\&= \frac{16a^2x^2y^2 + 4b^2(x^2-y^2+1)^2 + 16c^2y^2}{(1+x^2+y^2)^4}.
\end{split}
\end{equation}
Using Eq.~\eqref{eqt:mu_metric} and \eqref{eqt:E}--\eqref{eqt:G}, the Beltrami coefficient $\mu_{(P^{N}_{a,b,c})^{-1}}$ can be explicitly calculated at every point $(x,y)$ on the plane. In practice, the computation of the LBS method requires the Beltrami coefficient on the triangular faces. We can define $\mu_{(P^{N}_{a,b,c})^{-1}}$ on every triangular face $T = [p_1,p_2,p_3]$ on the plane as
\begin{equation}
    \mu_{(P^{N}_{a,b,c})^{-1}}(T) = \frac{\mu_{(P^{N}_{a,b,c})^{-1}}(p_1)+\mu_{(P^{N}_{a,b,c})^{-1}}(p_2)+\mu_{(P^{N}_{a,b,c})^{-1}}(p_3)}{3}, 
\end{equation}
where $p_1 = (x_1, y_1), p_2 = (x_2, y_2), p_3 = (x_3, y_3)$ are the three vertices of $T$.

Now, since $ \mu_{\psi} \circ \psi^{-1} = - \left( {\psi^{-1}_z} / {\overline{\psi^{-1}_{z}}} \right) \mu_{\psi^{-1}}$, we have
\begin{equation}
\begin{split}
    \mu_{\psi^{-1}} + \frac{\overline{\psi^{-1}_{z}}}{\psi^{-1}_z} \left(\mu_{(P^{N}_{a,b,c})^{-1}} \circ \psi^{-1}\right) 
    &= \mu_{\psi^{-1}} + \frac{\overline{\psi^{-1}_{z}}}{\psi^{-1}_z} (\mu_{\psi} \circ \psi^{-1}) \\
    &= \mu_{\psi^{-1}} + \frac{\overline{\psi^{-1}_{z}}}{\psi^{-1}_z}  \left(-\left( {\psi^{-1}_z} / {\overline{\psi^{-1}_{z}}} \right)  \mu_{\psi^{-1}} \right) \\
    &= \mu_{\psi^{-1}}  - \mu_{\psi^{-1}} \\
    &= 0.
\end{split}
\end{equation}
Hence, by the composition formula in Eq.~\eqref{eqt:composition}, we have
\begin{equation}
\begin{split}
    \mu_{(P^{N}_{a,b,c})^{-1} \circ \psi^{-1}} &= \frac{\mu_{\psi^{-1}} + \frac{\overline{\psi^{-1}_{z}}}{\psi^{-1}_z} (\mu_{(P^{N}_{a,b,c})^{-1}} \circ \psi^{-1})}{1+\frac{\overline{\psi^{-1}_{z}}}{\psi^{-1}_z} \overline{\mu_{\psi^{-1}}} (\mu_{(P^{N}_{a,b,c})^{-1}} \circ \psi^{-1})} \\
    &= \frac{0}{1+\frac{\overline{\psi^{-1}_{z}}}{\psi^{-1}_z} \overline{\mu_{\psi^{-1}}} (\mu_{(P^{N}_{a,b,c})^{-1}} \circ \psi^{-1})} = 0.
    \end{split}
\end{equation}
Therefore, $(P^{N}_{a,b,c})^{-1} \circ \psi^{-1}$ is conformal. In other words, $(P^{N}_{a,b,c})^{-1} \circ \psi^{-1}$ gives a conformal inverse projection from the extended complex plane $\overline{\mathbb{C}}$ to the ellipsoid $\mathcal{E}_{a,b,c}$.

Combining all the above-mentioned procedures, the final ellipsoidal conformal parameterization is given by:
\begin{equation} \label{eqt:final}
\begin{aligned}
    f &= (P_{a,b,c}^N)^{-1} \circ \psi^{-1} \circ k \circ g \circ {P^N} \circ \varphi.
\end{aligned}
\end{equation}
It is noteworthy that the stenographic projection ${P^N}$, the M\"obius transformation $g$, the rescaling $k$, and the inverse ellipsoidal stereographic projection $(P_{a,b,c}^N)^{-1}$ are naturally bijective. Also, the bijectivity of the initial parameterization $\varphi$ and the quasi-conformal map $\psi^{-1}$ is guaranteed by quasi-conformal theory. Therefore, the overall mapping $f$ is also bijective. The proposed fast ellipsoidal conformal map (FECM) algorithm is summarized in Algorithm~\ref{alg:fecm}.

\begin{algorithm}[h]
\KwIn{A genus-0 closed triangulated surface $\mathcal{M}$ and the target elliptic radii $a,b,c$.}
\KwOut{An ellipsoidal conformal parameterization $f:\mathcal{M} \to \mathcal{E}_{a,b,c}$.}
\BlankLine
Compute a spherical conformal parameterizations $\varphi: \mathcal{M} \to \mathbb{S}^2$ using~\cite{choi2015flash}\;
Apply the stereographic projection ${P^N}:\mathbb{S}^2 \to \overline{\mathbb{C}}$ using Eq.~\eqref{eqt:stereographic} to map the spherical parameterization onto the plane\;
Apply a M\"obius transformation $g:  \overline{\mathbb{C}} \to  \overline{\mathbb{C}}$ using Eq.~\eqref{eqt:mobius_alignment}\;
Apply the balancing scheme using the scaling factor $k$ in Eq.~\eqref{eqt:k}\;
Compute a quasi-conformal map $\psi: \overline{\mathbb{C}} \to \overline{\mathbb{C}}$ obtained using Eq.~\eqref{eqt:psi}\;
Compose the inverse ellipsoidal stereographic projection $(P_{a,b,c}^N)^{-1}$ in Eq.~\eqref{eqt:inverse_ellipsoidal_stereographic} with $\psi^{-1}$ to form a conformal inverse projection\;
The final ellipsoidal conformal parameterization is given by
$f = (P_{a,b,c}^N)^{-1} \circ \psi^{-1}  \circ k \circ g \circ {P^N} \circ \varphi$\;
\caption{Fast ellipsoidal conformal map (FECM) of genus-0 closed surfaces.}
\label{alg:fecm}
\end{algorithm}

\subsection{Optimizing the shape of the ellipsoidal domain}\label{sect:optimization}
In the above FECM algorithm, the shape of the target ellipsoid (i.e. the values of the elliptic radii $a,b,c$) is prescribed as an input. However, in some situations, it may be more desirable to determine the values of $a,b,c$ automatically. In this section, we develop an extension of Algorithm~\ref{alg:fecm} with the optimal values of $a,b,c$ automatically determined throughout the computation.

\subsubsection{Initializing the elliptic radii defined using a bounding box}
A simple way to obtain a suitable initial guess of the elliptic radii $(a,b,c) = (a_0, b_0, c_0)$ is to consider a bounding box of the given surface $\mathcal{M}$. Specifically, we first zero-center $\mathcal{M}$ and then rotate it to align it with the coordinate axes. We can then initialize the shape of the ellipsoid using the bounding box with side lengths ${x_{\max}-x_{\min}}$, ${y_{\max}-y_{\min}}$, and ${z_{\max}-z_{\min}}$, where $x_{\max}, y_{\max}, z_{\max}$ are the maximum $x, y, z$-coordinates of the surface and $x_{\min}, y_{\min}, z_{\min}$ are the minimum $x, y, z$-coordinates. To remove the effect of size in the subsequent optimization process, we can further normalize $a_0, b_0, c_0$ as follows:
\begin{equation}\label{eqt:abc0}
    a_0 = \frac{x_{\max}-x_{\min}}{K}, \ \ b_0 = \frac{y_{\max}-y_{\min}}{K}, \ \ c_0 = \frac{z_{\max}-z_{\min}}{K},
\end{equation}
where $K = \frac{1}{3}\left((x_{\max}-x_{\min})+(y_{\max}-y_{\min})+(z_{\max}-z_{\min})\right)$ is the average value of the bounding box side lengths.

\subsubsection{Optimizing the elliptic radii based on area distortion}
Note that by the properties of the mappings involved in Algorithm~\ref{alg:fecm}, the resulting ellipsoidal parameterization is conformal for any input elliptic radii theoretically. However, the parameterization is not necessarily area-preserving in general. It is therefore natural to search for an optimal set of elliptic radii such that the area distortion of the parameterization is minimized. Here, we consider using the logged area ratio~\cite{choi2020parallelizable} to assess the area distortion of the parameterization $f$:
\begin{equation} \label{eqt:d_area}
    d_{\text{area}}(T) = \log\left(\frac{\text{Area}(f(T))/\sum_{T'\in \mathcal{F}} \text{Area}(f(T'))}{\text{Area}(T)/\sum_{T'\in \mathcal{F}} \text{Area}(T')}\right),
\end{equation}
where $T \in \mathcal{F}$ is a triangular face in the given surface mesh, $f(T)$ is the corresponding triangular face in the parameterization $f$, and the two summations $\sum_{T'\in \mathcal{F}} \text{Area}(f(T'))$ and $\sum_{T'\in \mathcal{F}} \text{Area}(T')$ are normalization factors. It is easy to see that $ d_{\text{area}}=0$ if and only if the area of every face is perfectly preserved under the mapping $f$. We can then define the area distortion energy as
\begin{equation}
    E_{\text{area}} = \frac{1}{|\mathcal{F}|} \sum_{T \in \mathcal{F}} (d_{\text{area}}(T))^2
\end{equation}
and minimize it with respect to $a,b,c$. 

More specifically, we first follow the procedure in Algorithm~\ref{alg:fecm} and compute a parameterization $f_0 = (P_{a_0,b_0,c_0}^N)^{-1} \circ \psi^{-1}  \circ k \circ g \circ {P^N} \circ \varphi$ using the initial elliptic radii $(a,b,c) = (a_0, b_0, c_0)$. We then update the elliptic radii iteratively based on $-\nabla E_{\text{area}}$. Here, note that the initial spherical conformal parameterization $\varphi$, the stereographic projection ${P^N}$, the M\"obius transformation $g$, and the rescaling $k$ are all independent of $(a,b,c)$, and so we do not need to repeat their computation. Once we obtain the new mappings in the remaining quasi-conformal composition step based on the updated elliptic radii $(a_n, b_n, c_n)$, we have a new parameterization $f_n = (P_{a_n,b_n,c_n}^N)^{-1} \circ \psi^{-1}  \circ k \circ g \circ {P^N} \circ \varphi$. We can then replace $n$ with $n+1$ and repeat the above process until convergence.

To obtain the gradient $-\nabla E_{\text{area}}$, we first note that $\sum_{T'\in \mathcal{F}} \text{Area}(f(T'))$ should be close to the total surface area of the target ellipsoid $\mathcal{E}_{a,b,c}$, which can be approximated using the Knud Thomsen's formula (see also~\cite{klamkin1971elementary}):
\begin{equation}
A \approx 4\pi \left(\frac{a^p b^p + b^p c^p + c^p a^p}{3} \right)^{1/p},
\end{equation}
where $p \approx 1.6075$. It is easy to see that 
\begin{align} 
    \frac{\partial A}{\partial a} &=  \frac{4\pi}{3^{1/p}} a^{p-1}(b^p+c^p) \left({a^p b^p + b^p c^p + c^p a^p} \right)^{\frac{1}{p} - 1},\\
    \frac{\partial A}{\partial b} &=  \frac{4\pi}{3^{1/p}}  b^{p-1}(a^p+c^p) \left({a^p b^p + b^p c^p + c^p a^p} \right)^{\frac{1}{p} - 1},\\
  \frac{\partial A}{\partial c} &= \frac{4\pi}{3^{1/p}}  c^{p-1}(a^p+b^p) \left({a^p b^p + b^p c^p + c^p a^p} \right)^{\frac{1}{p} - 1}.
\end{align}
Second, note that the energy $E_{\text{area}}$ can be simplified as follows:
\begin{equation}
\begin{split}
    E_{\text{area}} &=  \frac{1}{|\mathcal{F}|} \sum_{T \in \mathcal{F}} \left(\log\left(\frac{\text{Area}\left((P_{a,b,c}^N)^{-1} ([z_i, z_j, z_k])\right)/A}{\text{Area}(T)/\sum_{T'\in \mathcal{F}} \text{Area}(T')}\right)\right)^2,
\end{split}
\end{equation}
where $[z_i, z_j, z_k]$ is the triangle corresponding to $T$ right before the final inverse ellipsoidal stereographic projection $(P_{a,b,c}^N)^{-1}$. From Eq.~\eqref{eqt:inverse_ellipsoidal_stereographic}, we have
\begin{align}
\frac{\partial (P_{a,b,c}^N)^{-1}(z)}{\partial a} &= \left(\frac{2 \text{Re}{(z)}}{1+|z|^2}, 0, 0\right),\\
\frac{\partial (P_{a,b,c}^N)^{-1}(z)}{\partial b} &= \left(0, \frac{2 \text{Im}{(z)}}{1+|z|^2}, 0\right),\\
\frac{\partial (P_{a,b,c}^N)^{-1}(z)}{\partial c} &= \left(0, 0, \frac{-1 +|z|^2}{1+|z|^2}\right).
\end{align}
Now, note that the area of the triangle $(P_{a,b,c}^N)^{-1} ([z_i, z_j, z_k])$ can be expressed in terms of the lengths of its edges $\left[(P_{a,b,c}^N)^{-1}(z_i),(P_{a,b,c}^N)^{-1}(z_j)\right]$, $\left[(P_{a,b,c}^N)^{-1}(z_j),(P_{a,b,c}^N)^{-1}(z_k)\right]$, $\left[(P_{a,b,c}^N)^{-1}(z_k),(P_{a,b,c}^N)^{-1}(z_i)\right]$, and hence we can easily get the partial derivatives of the area term with respect to $a,b,c$. Based on the above results, we can compute $\nabla E_{\text{area}}$ and perform gradient descent with:
\begin{equation}
    (a_{n+1}, b_{n+1}, c_{n+1}) = (a_{n}, b_{n}, c_{n}) - \gamma \nabla E_{\text{area}}(a_{n}, b_{n}, c_{n}),
\end{equation}
where $\gamma$ is the step size. We can then repeat the mapping procedure until the result converges, thereby getting an ellipsoidal conformal parameterization with optimized elliptic radii.

\section{Fast ellipsoidal quasi-conformal map (FEQCM)} \label{sect:main_qc}
Besides conformal parameterization, 
some prior spherical mapping works are capable of achieving quasi-conformal parameterizations satisfying different prescribed constraints~\cite{choi2015flash,choi2016fast}. It is natural to ask whether we can obtain ellipsoidal quasi-conformal parameterizations for genus-0 closed surfaces analogously.

Here, we develop a method for computing ellipsoidal quasi-conformal parameterizations with prescribed landmark constraints. Specifically, let $\mathcal{M}$ be a genus-0 closed surface. Our goal is to find an optimal ellipsoidal quasi-conformal parameterization $f: \mathcal{M} \to \mathcal{E}_{a,b,c}$ with a set of prescribed landmark constraints $\{p_i\}_{i=1}^m \subset \mathcal{M} \leftrightarrow \{q_i\}_{i=1}^m \subset \mathcal{E}_{a,b,c}$ such that the mapped positions $f(p_i)$ are close to the target positions $q_i$ for all $i=1,2,\dots, m$.

To begin, we follow the first few steps in the above-mentioned FECM method to apply the method in~\cite{choi2015flash} to get a spherical conformal parameterizations $\varphi: \mathcal{M} \to \mathbb{S}^2$, followed by the stereographic projection ${P^N}:\mathbb{S}^2 \to \overline{\mathbb{C}}$ using Eq.~\eqref{eqt:stereographic}, a M\"obius transformation $g: \overline{\mathbb{C}} \to  \overline{\mathbb{C}}$ using Eq.~\eqref{eqt:mobius_alignment}, and the balancing scheme with the scaling factor $k$ in Eq.~\eqref{eqt:k}.

Now, to obtain a quasi-conformal parameterization satisfying the given landmark matching conditions, we introduce a new step of computing a planar quasi-conformal map $\Phi: \overline{\mathbb{C}} \to \overline{\mathbb{C}}$ in the ellipsoidal parameterization process. More specifically, here we apply the FLASH method~\cite{choi2015flash} to solve for a landmark-aligned optimized harmonic map $\Phi$ with
\begin{equation} \label{eqt:flash}
    \Phi = \text{argmin} \left(\int |\nabla \phi|^2 + \lambda \sum_{i=1}^m |\phi(\widetilde{p}_i) - \widetilde{q}_i|^2\right),
\end{equation}
where $\lambda > 0$ is a weighting factor, and $\{\widetilde{p}_i\}_{i=1}^m, \{\widetilde{q}_i\}_{i=1}^m$ are the landmarks on $\overline{\mathbb{C}}$ associated with the given landmarks on $\mathcal{M}$ and $\mathcal{E}_{a,b,c}$ respectively. In particular, by changing the value of $\lambda$, we can achieve a balance between the landmark mismatch and the conformal distortion. If a smaller $\lambda$ is used, the mapping will be closer to conformal while the landmark mismatch will be larger. If a larger $\lambda$ is used, the mapping will become less conformal but the landmark mismatch will be reduced. The method then further ensures the bijectivity of the mapping $\Phi$ via an iterative process of modifying the Beltrami coefficient of $\Phi$ and reconstructing a new quasi-conformal map using the LBS method. 
    
Once we have obtained a quasi-conformal map $\Phi$ from the above-mentioned process, we can perform the remaining quasi-conformal composition step as in Algorithm~\ref{alg:fecm}. Finally, the resulting ellipsoidal quasi-conformal parameterization is given by:
\begin{equation} \label{eqt:final_qc}
\begin{aligned}
    f &= (P_{a,b,c}^N)^{-1} \circ {\psi^{-1}} \circ \Phi \circ k \circ g \circ {P^N} \circ \varphi.
\end{aligned}
\end{equation}
Analogous to the FECM method, since the bijectivity of each mapping is guaranteed by quasi-conformal theory, the overall parameterization $f$ is also bijective. The proposed fast ellipsoidal quasi-conformal map (FEQCM) algorithm is summarized in Algorithm~\ref{alg:feqcm}.

\begin{algorithm}[h]
\KwIn{A genus-0 closed triangulated surface $\mathcal{M}$, the target elliptic radii $a,b,c$, a set of prescribed landmark pairs $\{p_i\}_{i=1}^m \leftrightarrow \{q_i\}_{i=1}^m$, and a weighting factor $\lambda > 0$.}
\KwOut{A landmark-aligned ellipsoidal quasi-conformal map $f:\mathcal{M} \to \mathcal{E}_{a,b,c}$.}
\BlankLine

Compute a spherical conformal parameterizations $\varphi: \mathcal{M} \to \mathbb{S}^2$ using~\cite{choi2015flash}\;
Apply the stereographic projection ${P^N}:\mathbb{S}^2 \to \overline{\mathbb{C}}$ using Eq.~\eqref{eqt:stereographic} to map the spherical parameterization onto the plane\;
Apply a M\"obius transformation $g:  \overline{\mathbb{C}} \to  \overline{\mathbb{C}}$ using Eq.~\eqref{eqt:mobius_alignment}\;
Apply the balancing scheme using the scaling factor $k$ in Eq.~\eqref{eqt:k}\;
Compute a landmark-aligned quasi-conformal mapping $\Phi: \overline{\mathbb{C}} \to \overline{\mathbb{C}}$ using the FLASH method~\cite{choi2015flash} based on the landmark pairs $\{\widetilde{p}_i\}_{i=1}^m \leftrightarrow \{\widetilde{q}_i\}_{i=1}^m$ on $\overline{\mathbb{C}}$\;
Compute a quasi-conformal map $\psi: \overline{\mathbb{C}} \to \overline{\mathbb{C}}$ obtained using Eq.~\eqref{eqt:psi}\;
Compose the inverse ellipsoidal stereographic projection $(P_{a,b,c}^N)^{-1}$ in Eq.~\eqref{eqt:inverse_ellipsoidal_stereographic} with $\psi^{-1}$ to form a conformal inverse projection\;
The final ellipsoidal quasi-conformal parameterization is given by 
$f = (P_{a,b,c}^N)^{-1} \circ {\psi^{-1}} \circ \Phi \circ k \circ g \circ {P^N} \circ \varphi$\;
\caption{Fast ellipsoidal quasi-conformal map (FEQCM) of genus-0 closed surfaces.}
\label{alg:feqcm}
\end{algorithm}

\section{Experiments}\label{sect:experiments}
The proposed algorithms are implemented in MATLAB. Genus-0 mesh models are adapted from online mesh repositories~\cite{repositories} for assessing the performance of our proposed methods. All experiments are performed on a desktop computer with an Intel(R) Core(TM) i9-12900 2.40GHz processor and 32GB RAM.

\begin{figure}[t!]
    \centering
    \includegraphics[width=0.8\textwidth]{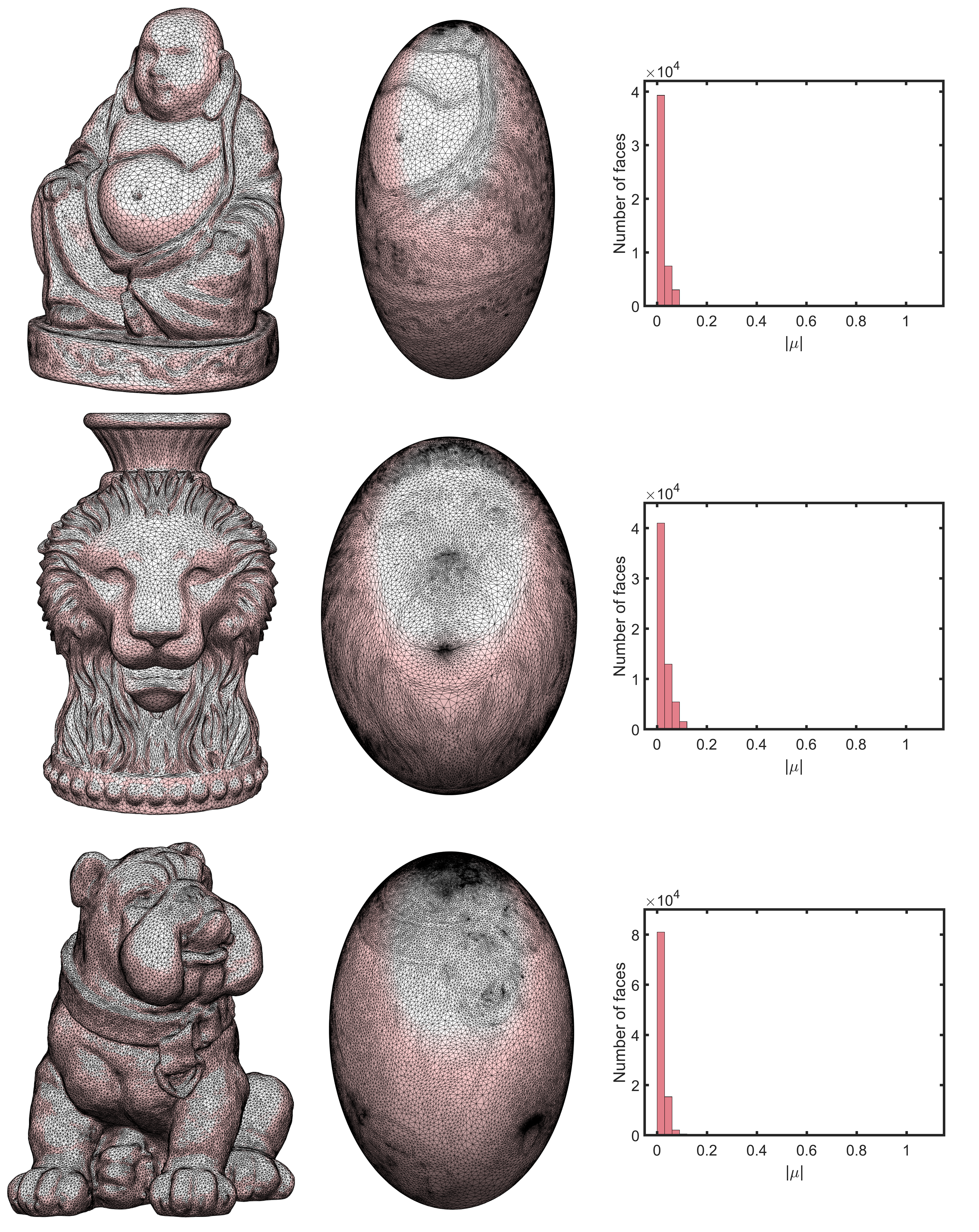}
    \caption{Examples of ellipsoidal conformal parameterization results obtained by our FECM method. Each row shows one example. (Left) The input genus-0 closed surface. (Middle) The resulting ellipsoidal conformal parameterization. (Right) The histogram of the norm of the Beltrami coefficient $|\mu|$.}
    \label{fig:results_fecm}
\end{figure}

\subsection{Ellipsoidal conformal parameterization}
We start by considering several genus-0 closed surfaces and computing the ellipsoidal conformal parameterizations using our proposed FECM method. As shown in Fig.~\ref{fig:results_fecm}, the local geometric features of the input surfaces are well-preserved under our ellipsoidal conformal parameterizations. To examine the conformality of the parameterizations, we compute the norm of the Beltrami coefficient $|\mu|$ for every triangular face in the ellipsoidal parameterizations. For all examples, the histogram of $|\mu|$ shows that $|\mu|$ is highly concentrated at 0, which indicates that the conformal distortion of the parameterization is very low. 

\begin{figure}[t!]
    \centering
    \includegraphics[width=\textwidth]{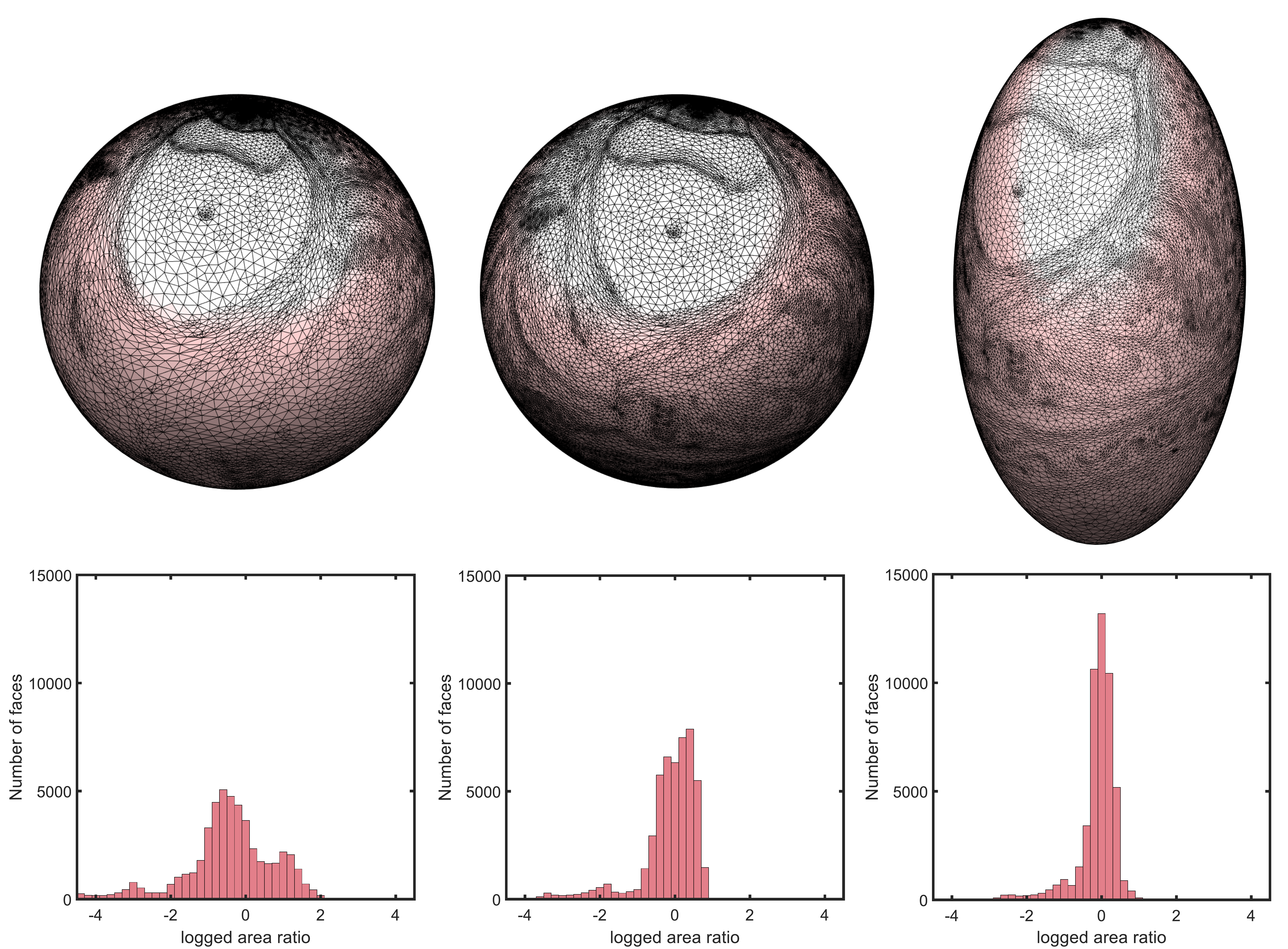}
    \caption{A comparison between spherical conformal parameterizations and the ellipsoidal conformal parameterization obtained by our FECM method. The Buddha model in Fig.~\ref{fig:results_fecm} is used as input, and each column shows one set of parameterization results. (Left) The spherical conformal parameterization obtained by~\cite{choi2015flash} and the histogram of the logged area ratio $d_{\text{area}}$. (Middle) The spherical conformal parameterization obtained by~\cite{choi2015flash} together with the M\"obius area correction scheme~\cite{choi2020parallelizable} and the histogram of $d_{\text{area}}$. (Right) The ellipsoidal conformal parameterization obtained by our proposed method and the histogram of $d_{\text{area}}$.}
    \label{fig:results_fecm_comparison}
\end{figure}

\begin{table}[t!]
\centering
\resizebox{\textwidth}{!}{
    \begin{tabular}{|C{25mm}|c|c|c|c|c|c|c|c|}
    \hline
    \textbf{Surface} & \textbf{\# Faces} & \textbf{Method} & \textbf{Time (s)} & \textbf{Mean($|\mu|$)} & \textbf{SD($|\mu|$)} & \textbf{Mean($|d_{\text{area}}|$)} & \textbf{SD($|d_{\text{area}}|$)} & \textbf{\# fold-overs}\\ \hline
    
    \multirow{3}{*}{Hippocampus} & \multirow{3}{*}{12K} 
    & \cite{choi2015flash} & 0.05 & 0.03 & 0.02 & 2.47 & 2.09 & 0\\ 
    & & \cite{choi2015flash}+\cite{choi2020parallelizable} & 0.34 & 0.03 & 0.02 & 2.43 & 1.77 & 0\\ 
    & & Ours $(1,1,2.3)$ & 0.08 & 0.03 & 0.02 & 1.32 & 1.51 & 0\\ \hline
    
    \multirow{3}{*}{Buddha} & \multirow{3}{*}{50K} 
    & \cite{choi2015flash} & 0.17 & 0.02 & 0.02 & 0.95 & 0.85 & 0\\
    & & \cite{choi2015flash}+\cite{choi2020parallelizable} & 2.14 & 0.02 & 0.02 & 0.51 & 0.58 & 0\\
    & & Ours $(1, 0.8, 1.5)$ & 0.38 & 0.02 & 0.02 & 0.32 & 0.41 & 0\\ \hline 
    
    \multirow{3}{*}{Bimba} & \multirow{3}{*}{50K} 
    & \cite{choi2015flash} & 0.17 & 0.02 & 0.02 & 2.45 & 2.45 & 0\\ 
    & & \cite{choi2015flash}+\cite{choi2020parallelizable} & 1.88 & 0.02 & 0.02 & 1.56 & 1.17 & 0 \\ 
    & & Ours $(1,1.05,1.5)$ & {0.38} & 0.02 & 0.02 & 1.54 & 0.96 &  0\\ \hline
    
    \multirow{3}{*}{Lion Vase} & \multirow{3}{*}{61K} 
    & \cite{choi2015flash} & 0.21 & 0.03 & 0.03 & 1.25 & 1.20 & 0\\ 
    & & \cite{choi2015flash}+\cite{choi2020parallelizable} & 2.52 & 0.03 & 0.03 & 0.79 & 0.78 & 0\\ 
    & & Ours $(1,1,1.4)$ & {0.47} & 0.03 & 0.03 & 0.66 & 0.77 & 0\\ \hline

    \multirow{3}{*}{Moai} & \multirow{3}{*}{96K} 
    & \cite{choi2015flash} & 0.37 & 0.02 & 0.03 & 2.21 & 1.93 & 0\\ 
    & & \cite{choi2015flash}+\cite{choi2020parallelizable} & 5.42 & 0.02 & 0.03 & 1.44 & 1.31 & 0\\ 
    & & Ours $(1,1,2)$ & {0.85} & 0.02 & 0.03 & 0.96 & 1.09 & 0\\ \hline

    \multirow{3}{*}{Bulldog} & \multirow{3}{*}{100K} 
    & \cite{choi2015flash} & 0.34 & 0.02 & 0.03 & 1.38 & 1.15 & 0\\ 
    & & \cite{choi2015flash}+\cite{choi2020parallelizable} & 3.70 & 0.02 & 0.03 & 1.08 & 0.91 & 0\\ 
    & & Ours $(1,1.1,1.5)$ & {0.78} & 0.02 & 0.03 & 1.00 & 0.88 & 0\\ \hline
    
    \multirow{3}{*}{Nefertiti} & \multirow{3}{*}{100K} 
    & \cite{choi2015flash} & 0.36 & 0.02 & 0.02 & 5.87 & 3.37 & 0\\ 
    & & \cite{choi2015flash}+\cite{choi2020parallelizable} & 5.12 & 0.02 & 0.02 & 2.12 & 2.59 & 0\\ 
    & & Ours $(1,1,2.4)$ & {0.82} & 0.02 & 0.02 & 1.73 & 1.18 & 0\\ \hline

    \end{tabular}
    }
    \caption{Comparison between the spherical conformal parameterization method in~\cite{choi2015flash}, the method in~\cite{choi2015flash} together with the M\"obius area correction scheme in~\cite{choi2020parallelizable}, and our proposed ellipsoidal conformal parameterization method. For each method, we record the computational time (in seconds), the mean and standard deviation of the norm of the Beltrami coefficient $|\mu|$, the mean and standard deviation of the norm of the logged area ratio $|d_{\text{area}}|$, and the number of mesh fold-overs. The elliptic radii $(a,b,c)$ used in our method are also provided.}
    \label{tab:comparison_spherical}
\end{table}

In Fig.~\ref{fig:results_fecm_comparison}, we compare our FECM method with the fast spherical conformal parameterization method in~\cite{choi2015flash} and the method in~\cite{choi2015flash} together with the M\"obius area correction scheme in a more recent work~\cite{choi2020parallelizable}, which searches for an optimal M\"obius transformation to further reduce the area distortion of the spherical conformal parameterization. Specifically, while all of the above-mentioned methods are conformal, the area distortion of them can be different. Here, we quantify the area distortion of the parameterization by considering the logged area ratio { in Eq.~\eqref{eqt:d_area}. It} can be observed from the area distortion histograms in Fig.~\ref{fig:results_fecm_comparison} that the spherical conformal parameterization by~\cite{choi2015flash} leads to very large area distortion, and the M\"obius area correction scheme~\cite{choi2020parallelizable} can only improve the area distortion of the spherical parameterization to some extent. By contrast, using our proposed FECM method, the area distortion can be significantly reduced.

For a more quantitative analysis, in Table~\ref{tab:comparison_spherical} we compare our proposed FECM method with the above-mentioned spherical conformal parameterization methods in terms of efficiency, conformality, area distortion, and bijectivity by evaluating the computational time, the norm of the Beltrami coefficient $|\mu|$, the norm of the logged area ratio $|d_{\text{area}}|$, and the number of mesh fold-overs, respectively (see also Fig.~\ref{fig:illustration_fecm} and Fig.~\ref{fig:results_fecm} for the surface models and the ellipsoidal parameterizations). It can be observed that while the method in~\cite{choi2015flash} requires the lowest computational cost, it results in a very high area distortion, which can be explained by the fact that the sphere may not be the best parameter domain for the given surfaces. By including an additional step of finding an optimal M\"obius transformation, the combination of \cite{choi2015flash} and \cite{choi2020parallelizable} may improve the spherical conformal parameterization and give a lower area distortion. However, the computation is significantly slower because of the additional optimization step. As our FECM method uses~\cite{choi2015flash} in the initial spherical parameterization step, it is natural that our method takes a longer computational time than the method~\cite{choi2015flash}. Nevertheless, we can see that our method is much faster than the M\"obius approach and is capable of further reducing the area distortion when compared with both of the above-mentioned approaches. Besides, the conformality and bijectivity of our method are comparable to the two spherical parameterization approaches. This can be explained by the fact that we have extra flexibility in controlling the geometry of the ellipsoidal parameterization domain, which allows us to further reduce the geometric distortion of the parameterizations.

\begin{table}[t]
\centering
\resizebox{\textwidth}{!}{
    \begin{tabular}{|C{25mm}|c|c|c|c|c|c|}
    \hline
    \textbf{Surface} & \textbf{\# Faces} & \textbf{Method} & \textbf{Mean($|\mu|$)} & \textbf{SD($|\mu|$)} & \textbf{Mean($|d_{\text{area}}|$)} & \textbf{SD($|d_{\text{area}}|$)}\\ \hline
    
    \multirow{3}{*}{Hippocampus} & \multirow{3}{*}{12K} 
    & \cite{choi2015flash} & 0.40 & 0.07 & 2.37 & 2.07\\ 
    & & \cite{choi2015flash}+\cite{choi2020parallelizable} & 0.34 & 0.06 & 2.30 & 1.73\\ 
    & & Ours & 0.03 & 0.02 & 1.32 & 1.51\\ \hline
    
    \multirow{3}{*}{Buddha} & \multirow{3}{*}{50K} 
    & \cite{choi2015flash} & 0.19 & 0.08 & 1.33 & 1.00\\
    & & \cite{choi2015flash}+\cite{choi2020parallelizable} & 0.19 & 0.07 & 0.42 & 0.53\\
    & & Ours & 0.02 & 0.02 & 0.32 & 0.41\\ \hline 
    
    \multirow{3}{*}{Bimba} & \multirow{3}{*}{50K} 
    & \cite{choi2015flash} & 0.08 & 0.06 & 3.07 & 2.85\\ 
    & & \cite{choi2015flash}+\cite{choi2020parallelizable} & 0.16 & 0.04 & 1.49 & 1.13 \\ 
    & & Ours & 0.02 & 0.02 & 1.54 & 0.96 \\ \hline
    
    \multirow{3}{*}{Lion Vase} & \multirow{3}{*}{61K} 
    & \cite{choi2015flash} & 0.10 & 0.05 & 2.34 & 1.66\\ 
    & & \cite{choi2015flash}+\cite{choi2020parallelizable} & 0.14 & 0.05 & 0.75 & 0.77\\ 
    & & Ours & 0.03 & 0.03 & 0.66 & 0.77\\ \hline

    \multirow{3}{*}{Moai} & \multirow{3}{*}{96K} 
    & \cite{choi2015flash} & 0.21 & 0.09 & 2.61 & 2.19\\ 
    & & \cite{choi2015flash}+\cite{choi2020parallelizable} & 0.30 & 0.07 & 1.35 & 1.25\\ 
    & & Ours  & 0.02 & 0.03 & 0.96 & 1.09\\ \hline

    \multirow{3}{*}{Bulldog} & \multirow{3}{*}{100K} 
    & \cite{choi2015flash}  & 0.11 & 0.06 & 1.24 & 1.00\\ 
    & & \cite{choi2015flash}+\cite{choi2020parallelizable} & 0.14 & 0.05 & 1.02 & 0.89\\ 
    & & Ours & 0.02 & 0.03 & 1.00 & 0.88\\ \hline
    
    \multirow{3}{*}{Nefertiti} & \multirow{3}{*}{100K} 
    & \cite{choi2015flash} & 0.14 & 0.12 & 6.72 & 3.75\\ 
    & & \cite{choi2015flash}+\cite{choi2020parallelizable} & 0.35 & 0.08 & 2.00 & 2.51\\ 
    & & Ours & 0.02 & 0.02 & 1.73 & 1.18\\ \hline
    
    \end{tabular}
    }
    \caption{Comparison between the ellipsoidal parameterizations obtained by our proposed method, an extension of the method in~\cite{choi2015flash}, and an extension of the method in~\cite{choi2015flash} together with~\cite{choi2020parallelizable}. Specifically, all stereographic projections and their inverses in \cite{choi2015flash} and~\cite{choi2020parallelizable} are replaced with the ellipsoidal stereographic projections and their inverses. For each example, the same ellipsoidal domain is used for producing the three parameterization results. For each method, we record the mean and standard deviation of the norm of the Beltrami coefficient $|\mu|$ and the mean and standard deviation of the norm of the logged area ratio $|d_{\text{area}}|$.}
    \label{tab:comparison_ellipsoidal}
\end{table}

In Table~\ref{tab:comparison_ellipsoidal}, we further compare the ellipsoidal parameterization results obtained by different approaches with the same ellipsoidal domain used. Specifically, here we consider a direct extension of~\cite{choi2015flash} and \cite{choi2020parallelizable} by replacing all stereographic projections and inverse stereographic projections in them with the ellipsoidal stereographic projections and their inverses to produce ellipsoidal parameterizations. We then assess the conformal and area distortions of the ellipsoidal parameterization results obtained by these approaches as well as our proposed FCEM method. It can be observed that even with the same ellipsoidal parameterization domain used, directly extending the two prior approaches will lead to large geometric distortions. This can be explained by the fact that the ellipsoidal stereographic projections are not conformal in general, and so conformal distortions will be induced throughout the algorithm in~\cite{choi2015flash}. Also, the M\"obius area correction scheme in~\cite{choi2020parallelizable} relies on the fact that the composition of the stereographic projection and the M\"obius transformations is conformal, which does not necessarily hold if the ellipsoidal stereographic projection is used instead. Therefore, while the extended M\"obius area correction scheme can reduce the area distortion, it may lead to an increase in the conformal distortion. This shows that both using an ellipsoidal domain and developing a suitable parameterization method for it are important for achieving desirable parameterization results. Altogether, the results suggest that our method is more advantageous than the prior conformal parameterization approaches for practical applications.

\begin{figure}[t!]
    \centering
    \includegraphics[width=\textwidth]{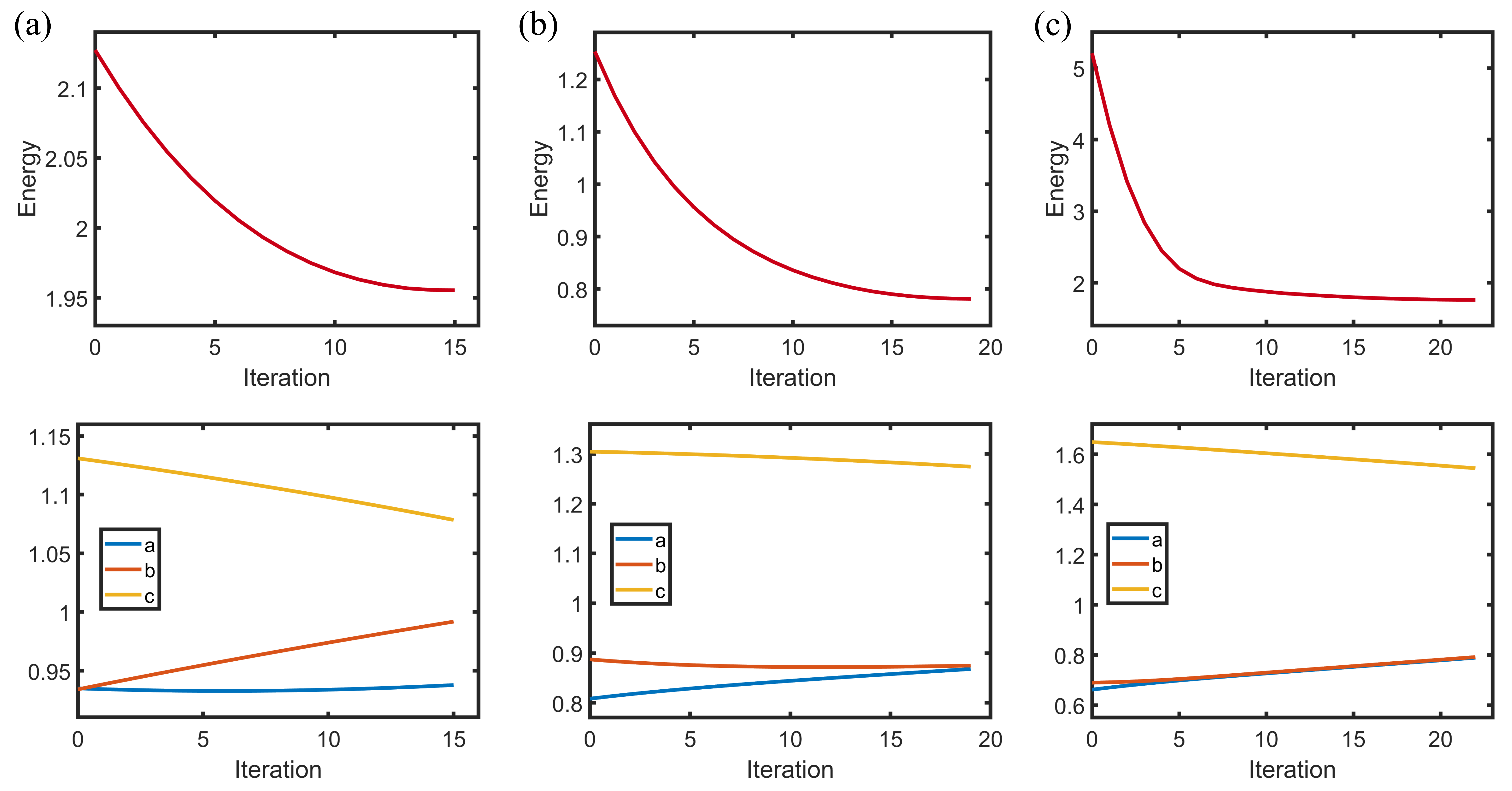}
    \caption{{Optimizing the elliptic radii for the ellipsoidal conformal parameterization. (a)~The result for the Bulldog model. (b) The result for the Lion Vase model. (c) The result for the Moai model. For each model, we show the change of the area distortion energy $E_{\text{area}}$ (top) and the elliptic radii $a,b,c$ (bottom) throughout the iterations.}}
    \label{fig:area_fecm}
\end{figure}

As discussed in Section~\ref{sect:optimization}, we can determine the optimal elliptic radii for the ellipsoidal conformal parameterization of a given surface via an energy minimization approach. To illustrate this idea, in Fig.~\ref{fig:area_fecm}(a) we consider finding the optimal radii for the ellipsoidal conformal parameterization of the Bulldog model. In this example, note that the initial $a$ and $b$ are almost identical, and the initial value of $c$ is relatively large. Throughout the optimization process, the values of $a,b,c$ are updated iteratively such that the area distortion energy $E_{\text{area}}$ decreases until convergence. In particular, it can be observed that the difference between $a$ and $b$ becomes prominent eventually and the value of $c$ decreases gradually. Fig.~\ref{fig:area_fecm}(b) shows another example of parameterizing the Lion Vase model, from which we again see that energy $E_{\text{area}}$ decreases gradually until convergence. In this example, the initial $a$ and $b$ are notably different, but they become close to each other throughout the iterations, yielding an approximately spheroidal shape in the final parameterization. Fig.~\ref{fig:area_fecm}(c) shows the result for the Moai model. Note that the initial ellipsoid is highly stretched, with the ratio of the maximum to minimum radii being around 2.5. Throughout the iterations, the ratio eventually decreases to around 1.9, and the energy $E_{\text{area}}$ is significantly reduced by over 60\%. The above experimental results show that the proposed optimization approach can effectively handle different surfaces and achieve ellipsoidal conformal parameterizations with the elliptic radii optimized based on the input surface geometry.

\begin{figure}[t!]
    \centering
    \includegraphics[width=0.8\textwidth]{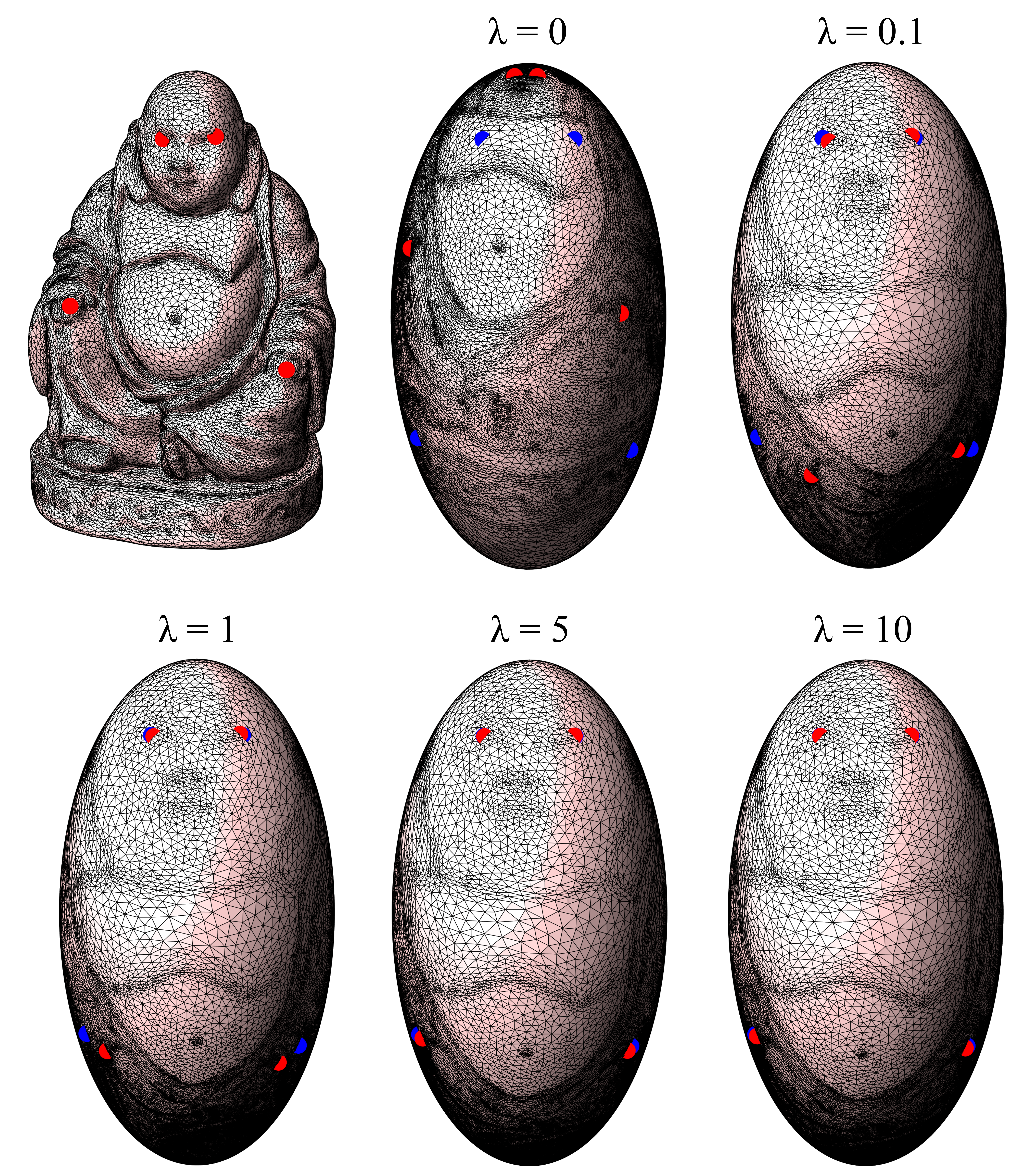}
    \caption{Landmark-aligned ellipsoidal quasi-conformal parameterizations of the Buddha model obtained by our FEQCM method. Top left: The input genus-0 closed surface with some labeled landmarks (red dots). Top middle: The ellipsoidal conformal parameterization with the labeled landmarks (red dots) and the prescribed target positions (blue dots). Top right and bottom left to right: The ellipsoidal quasi-conformal parameterization results with $\lambda = 0.1, 1, 5, 10$. }
    \label{fig:results_feqcm_buddha}
\end{figure}

\begin{figure}[t!]
    \centering
    \includegraphics[width=0.8\textwidth]{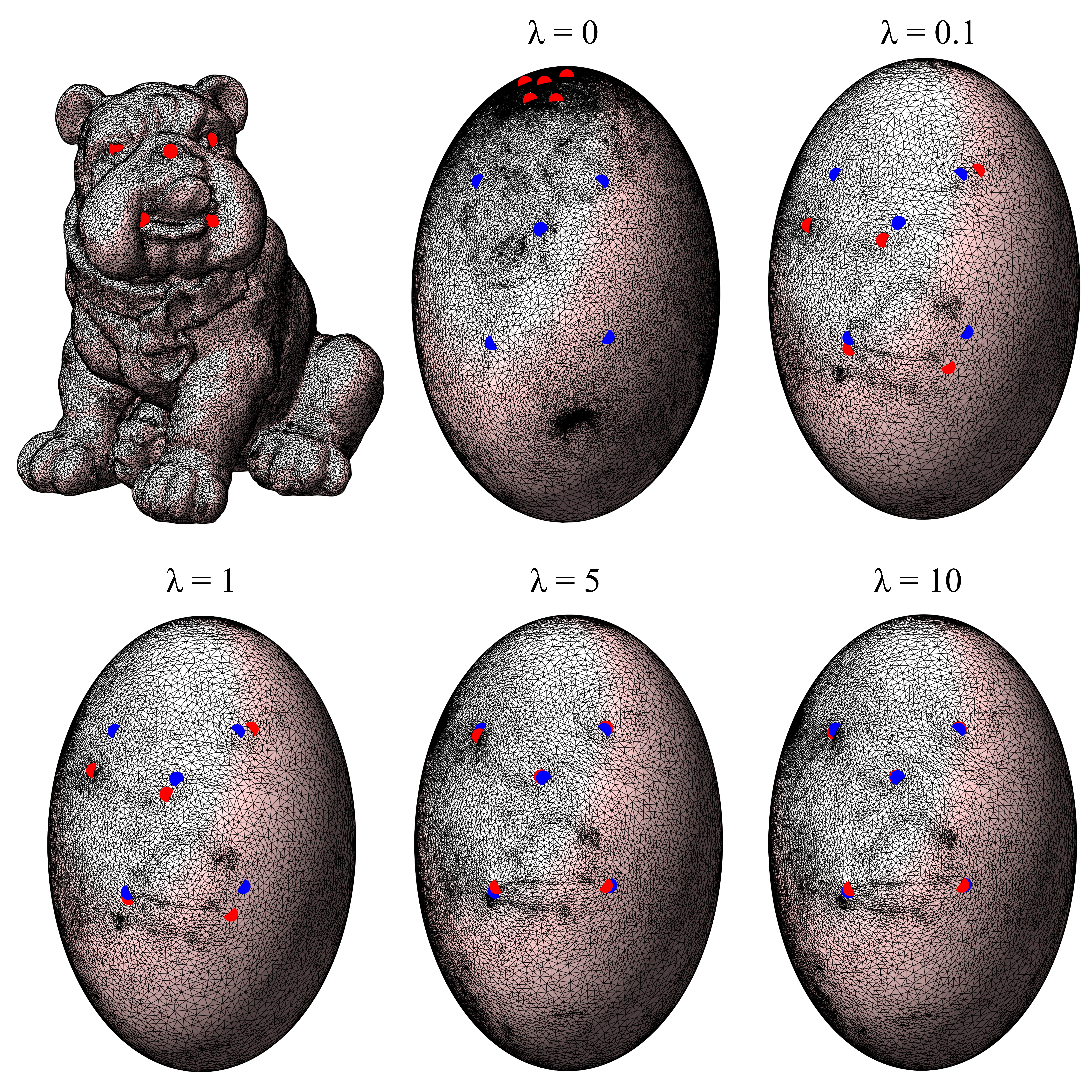}
    \caption{Landmark-aligned ellipsoidal quasi-conformal parameterizations of the Bulldog model obtained by our FEQCM method. Top left: The input genus-0 closed surface with some labeled landmarks (red dots). Top middle: The ellipsoidal conformal parameterization with the labeled landmarks (red dots) and the prescribed target positions (blue dots). Top right and bottom left to right: The ellipsoidal quasi-conformal parameterization results with $\lambda = 0.1, 1, 5, 10$. }
    \label{fig:results_feqcm_bulldog}
\end{figure}

\subsection{Ellipsoidal quasi-conformal parameterization}
After demonstrating the effectiveness of our proposed ellipsoidal conformal parameterization method, we assess the performance of our proposed FEQCM method for ellipsoidal quasi-conformal parameterizations.

Fig.~\ref{fig:results_feqcm_buddha} and Fig.~\ref{fig:results_feqcm_bulldog} show two sets of examples of ellipsoidal quasi-conformal parameterization results obtained by our FEQCM method with some labeled landmarks and prescribed target positions. It can be observed that the landmarks are largely mismatched in the ellipsoidal conformal parameterization obtained using the FECM algorithm (see the red and blue dots). By contrast, using the FEQCM method, the landmarks become well-aligned in the ellipsoidal quasi-conformal parameterization results. Specifically, with the weighting factor $\lambda$ of the landmark mismatch term in Eq.~\eqref{eqt:flash} increasing from $\lambda = 0.1$ to $\lambda = 10$, we can see that the difference between the labeled landmarks and the target positions becomes smaller and smaller. 

More quantitatively, we can assess the landmark mismatch error by considering the average of the 2-norm difference $\|f(p_i)-q_i\|_2$ between every pair of landmarks $(p_i, q_i)$ under the ellipsoidal parameterization $f$. From Table~\ref{tab:feqcm}, we can see that the landmark mismatch error decreases as the parameter $\lambda$ increases, while the conformal distortion will increase as a trade-off. Nevertheless, the bijectivity of the parameterizations remains unchanged. The results demonstrate the effectiveness of our proposed FEQCM method.

\begin{table}[t!]
    \centering
    \begin{tabular}{|c|c|c|c|c|}\hline
        \textbf{Surface} &  \textbf{Method} & \textbf{Mean($|\mu|$)} & \textbf{Mean($\|f(p_i)-q_i\|_2$)} & \textbf{\# fold-overs}\\ \hline
        
        \multirow{5}{*}{Buddha} & FECM & 0.02 & 0.76 & 0\\ 
        & FEQCM ($\lambda = 0.1$) & 0.02 & 0.17 & 0\\
        & FEQCM ($\lambda = 1$) & 0.04 & 0.11 & 0\\
        & FEQCM ($\lambda = 5$) & 0.08 & 0.02 & 0\\
        & FEQCM ($\lambda = 10$) & 0.09 & 0.01 & 0\\ \hline
        
        \multirow{5}{*}{Bulldog} & FECM & 0.02 & 1.21 & 0\\ 
        & FEQCM ($\lambda = 0.1$) & 0.02 & 0.19 & 0\\
        & FEQCM ($\lambda = 1$) & 0.03 & 0.15 & 0\\
        & FEQCM ($\lambda = 5$) & 0.04 & 0.03 & 0\\
        & FEQCM ($\lambda = 10$) & 0.05 & 0.01 & 0\\ \hline

    \end{tabular}
    
    \caption{The performance of the proposed fast ellipsoidal quasi-conformal mapping method. For each surface, we compute an ellipsoidal conformal parameterization using the FECM algorithm (Algorithm~\ref{alg:fecm}) and the ellipsoidal quasi-conformal parameterizations using the FEQCM algorithm (Algorithm~\ref{alg:feqcm}) with different weighting factor $\lambda$. For each mapping result, we assess the conformal distortion using the mean of the norm of the Beltrami coefficient $|\mu|$, the landmark mismatch error using the mean of the 2-norm difference $\|f(p_i)-q_i\|_2$ between every pair of landmarks $(p_i, q_i)$, and the bijectivity using the number of fold-overs in the mapping result.}
    \label{tab:feqcm}
\end{table}

\section{Discussion} \label{sect:conclusion}
In this paper, we have developed a novel method for the ellipsoidal conformal parameterization of genus-0 closed surfaces using a combination of conformal and quasi-conformal mappings. Using our proposed method, we can achieve comparable conformality and lower area distortion in the final parameterization results when compared to prior spherical parameterization approaches. We have further extended the method for computing ellipsoidal quasi-conformal parameterizations with prescribed landmark constraints. Experimental results on various surface models with different geometries have demonstrated the effectiveness of the proposed methods.

A possible future direction is to exploit parallelism for computing ellipsoidal parameterizations. Specifically, as different parallelizable methods for conformal parameterization~\cite{choi2020parallelizable} and quasi-conformal parameterization~\cite{zhu2022parallelizable} have been recently proposed, it is natural to consider extending these approaches for computing ellipsoidal conformal or quasi-conformal parameterization. Also, while the algorithms developed in this paper focus on triangle meshes, it may be possible to extend the computation for point clouds as in recent point cloud parameterization works~\cite{choi2016spherical,meng2016tempo,choi2022free}.

Besides, we plan to explore the possibility of combining the proposed ellipsoidal parameterization methods with density-equalizing mapping methods~\cite{choi2018density,choi2020area}, which produce area-based shape deformation based on prescribed density distribution. Specifically, we may be able to achieve ellipsoidal parameterizations with controllable area changes via such a combination of mapping methods.\\

\noindent \textbf{Acknowledgements} \ We thank Dr. Mahmoud Shaqfa (MIT) for helpful discussions. 

\bibliographystyle{ieeetr}
\bibliography{ellipsoidalbib.bib}

\end{document}